\documentclass[superscriptaddress, nofootinbib, amsmath, amssymb]{revtex4}
\usepackage{graphicx}
\usepackage[colorlinks=true,citecolor=blue,linkcolor=blue,urlcolor=blue]{hyperref}
\usepackage{color}
\begin{document}
\title{Degravitation and the relaxed Einstein equations}
\author{Alain Dirkes}
\email{dirkes@fias.uni-frankfurt.de}
\affiliation{Frankfurt Institute for Advanced Studies (FIAS), Goethe University Frankfurt,\\ Ruth-Moufang Str.1, Frankfurt am Main, D-60438, Germany}
\begin{abstract}
The general idea to modify Einstein's field equations by promoting Newton's constant $G$ to a covariant differential operator $G_\Lambda(\Box_g)$ was apparently outlined for the first time in \cite{Dvali1, Dvali2, Barvinsky1, Barvinsky2}. The modification itself originates from the quest of finding a mechanism which is able to {\it degravitate} the vacuum energy on cosmological scales. We present in this article a precise covariant coupling model which acts like a high-pass filter with a macroscopic distance filter scale $\sqrt{\Lambda}$. In the context of this particular theory of gravity we work out the effective relaxed Einstein equations as well as the effective 1.5 post-Newtonian total near-zone mass of a many body system. We observe that at any step of computation we recover in the limit of vanishing modification parameters the corresponding general relativistic result. 
\end{abstract}
\maketitle
\section{Introduction:}
In this chapter we will introduce the nonlocally modified Einstein field equations and outline how the vacuum energy is effectively degravitated on cosmological scales. In the second chapter we will briefly review the standard relaxed Einstein equations and their solutions in terms of a post-Newtonian expansion. In the third chapter we will work out the effective wave equation and provide a formal solution for a far away wave zone field point. Chapter four is devoted to the study of the nonlocally modified effective energy-momentum pseudotensor. In the penultimate chapter we combine the results worked out in the previous chapters in order to compute  the effective total near-zone mass. It should be noticed that each chapter has a separate appendix-section in which we present additional computational details.
\subsection{The nonlocally modified Einstein equations:}
It is well known that the essence of Einstein's field equations \cite{Einstein1} can be elegantly summarized by John A. Wheeler's famous words: matter tells spacetime how to curve and spacetime tells matter how to move. They relate indeed, by means of the Einstein curvature tensor $G^{\alpha\beta}$ and the total energy-momentum tensor $T^{\alpha\beta}$, the curvature of spacetime to the distribution of energy within spacetime, $G_{\alpha\beta}\,=\,\frac{8\pi}{c^4} \ G \ T_{\alpha\beta}$. Long before Albert Einstein published his theory of general relativity the relation between matter $\rho$ and the gravitational field $U$ had already been discovered and concisely summarized by the famous Poisson equation $\Delta U=-4\pi G \rho$. This law is purely phenomenological whereas Einstein's theory provides, via the concept of spacetime curvature, a deeper understanding of the true nature of gravity. One year after the final formulation of the theory of general relativity, Albert Einstein predicted the existence of gravitational radiation. He realized that the linearised weak-field equations admit solutions in the form of gravitational waves travelling at the speed of light. He also recognized that the direct experimental detection of these waves, which are generated by time variations of the mass quadrupole moment of the source, will be extremely challenging because of their remarkably small amplitude \cite{Einstein2, Einstein3}. However gravitational radiation has been detected indirectly since the mid seventies of the past century in the context of binary-systems \cite{Taylor1, Burgay1, Stairs1, Stairs2, Taylor2}. Precisely one century after Einstein's theoretical prediction, an international collaboration of scientists (LIGO Scientific Collaboration and Virgo Collaboration) reported the first direct observation of gravitational waves \cite{LIGO1,LIGO2,LIGO3}. The wave signal GW150914 was detected independently by the two LIGO detectors and its basic features point to the coalescence of two stellar black holes. Albeit the great experimental success of Einstein's theory, some issues, like the missing mass problem or the dark energy problem, the physical interpretation of black hole curvature singularities or the question of how a possible unification with quantum mechanics could be achieved, remain yet unsolved. In this regard many potentially viable alternative theories of gravity have been developed over the past decades. The literature on theories of modified gravity is rather long and we content ourselves here by providing an incomplete list of papers addressing this subject \cite{Will1, Esposito1, Clifton1, Tsujikawa1, Woodard1, BertiBuonannoWill}. In this article we aim to outline a particular model of a nonlocally modified theory of general relativity. The main difference between the standard field equations and the modified theory of gravity is that we promote the gravitational constant to a covariant differential operator,
\begin{equation}
\label{NonlocalEinstein}
G_{\alpha\beta}\,=\, \frac{8\pi}{c^4} \ G_\Lambda(\Box_g) \ T_{\alpha\beta},
\end{equation}
where $\Box_g=\nabla^{\alpha}\nabla_\alpha$ is the covariant d'Alembert operator and $\sqrt{\Lambda}$ is the scale at which infrared modifications become important. The general idea of a differential coupling was apparently formulated for the first time in \cite{Dvali1,Barvinsky1,Dvali2,Barvinsky2} in order to address the cosmological constant problem \cite{Weinberg1}. However the idea of a varying coupling constant of gravitation dates back to early works of Dirac \cite{Dirac1} and Jordan \cite{Jordan1, Jordan2}. Inspired by these considerations Brans and Dicke published in the early sixties a theory in which the gravitational constant is replaced by the reciprocal of a scalar field \cite{Brans1}. Further developments going in the same direction can be inferred from \cite{Narlikar1, Isern1, Uzan2}. Although we are going to present a purely bottom-up constructed model, it is worth mentioning that many theoretical approaches, such as models with extra dimensions, string theory or scalar–tensor models of quintessence \cite{Peebles1,Steinhardt1,Lykkas1} contain a built–in mechanism for a possible time variation of the couplings \cite{Dvali3, Dvali4, Dvali5,Parikh1,Damour1,Uzan1,Lykkas1}. The main difference between the standard general relativistic theory and our nonlocally modified theory is how the energy-momentum tensor source term is translated into spacetime curvature. In the usual theory of gravity this translation is assured by the gravitational coupling constant $G$, whereas in our modified approach the coupling between the energy source term and the gravitational field will be in the truest sense of the word more differentiated. The covariant d'Alembert operator is sensitive to the characteristic wavelength of the gravitating system under consideration $1/\sqrt{-\Box_g} \sim \lambda_c$. We will see that our precise model will be constructed in such a way that the long-distance modification is almost inessential for processes varying in spacetime faster than $1/\sqrt{\Lambda}$ and large for slower phenomena at wavelengths $\sim \sqrt{\Lambda}$ and larger. In this regard spatially extended processes varying very slowly in time, with a small characteristic frequency $\nu_c\sim 1/\lambda_c$, will produce a less stronger gravitational field than smaller fast moving objects like solar-system planets or even earth sized objects. The latter possess rather small characteristic wavelengths and will therefore couple to the gravitational field in the usual way. Cosmologically extended processes with a small characteristic frequency will effectively decouple from the gravitational field. In this regard it is of course understood that John Wheeler's famous statement about the mutual influence of matter and spacetime curvature remains essentially true, however the precise form of the coupling differs according to the dynamical nature of the gravitating object under consideration. Indeed promoting Newton's constant $G$ to a differential operator $G_{\Lambda}(\Box_g)$ allows for an interpolation between the Planckian value of the gravitational constant and its long distance magnitude \cite{Barvinsky1,Barvinsky2},
\begin{eqnarray*}
G_P>G_\Lambda(\Box_g)>G_{L}.
\end{eqnarray*}
Thus the differential operator acts like a high-pass filter with a macroscopic distance filter scale $\sqrt{\Lambda}$. In this way sources characterized by characteristic wavelengths much smaller than the filter scale  ($\lambda_c\ll\sqrt{\Lambda}$) pass undisturbed through the filter and gravitate normally, whereas sources characterized by wavelengths larger than the filter scale are effectively filtered out \cite{Dvali1,Dvali2}. In a more quantitative way we can see how this filter mechanism works by introducing the dimensionless parameter $z\,=\,-\Lambda \Box_g\sim \Lambda/\lambda_c^2$,
\begin{eqnarray*}
G(z)\rightarrow G, \ |z|\gg 1 \ (\lambda_c \ll 1),\quad \quad G(z)\rightarrow 0, \ \  |z|\ll 1 \ (\lambda_c \gg 1).
\end{eqnarray*}
For small and fast moving objects with large values of $|z|$ (small characteristic wavelengths) the covariant coupling operator will essentially reduce to Newton's constant $G$, whereas for slowly varying processes characterized by small values of $|z|$ (large characteristic wavelengts) the coupling will be much smaller.  Although the equations of motion $\eqref{NonlocalEinstein}$ are themselves generally covariant, they cannot, for nontrivial $G_\Lambda(\Box_g)$, be represented as a metric variational derivative of a diffeomorphism invariant action. The solution of this problem was suggested in \cite{Barvinsky1,Barvinsky2, Modesto1} by viewing equation $\eqref{NonlocalEinstein}$ only as a first, linear in the curvature, approximation for the correct equations of motion. Their covariant action can be constructed as a weak-field expansion in powers of the curvature with nonlocal coefficients. The nonlocally modified action $S_{NL}[g_{\mu\nu}]$ should be derived from the variational equation, 
\begin{equation}
\label{Variational}
\frac{\delta S_{NL}[g_{\mu\nu}]}{\delta g_{\mu\nu}(x)}\,=\,\frac{c^3}{16 \pi G_\Lambda(\Box_g)}\sqrt{-g}\ G^{\mu\nu}+\mathcal{O}[R^2_{\mu\nu}],
\end{equation}
where we remind that $G_{\mu\nu}=R_{\mu\nu}-\frac{1}{2}g_{\mu\nu}R$ is the Einstein tensor and $R_{\mu\nu}$ the Riemannian curvature tensor. In order to obtain the leading term of $S_{NL}$, the equation above can be functionally integrated with the aid of the covariant curvature expansion technique presented in \cite{Barvinsky1, Barvinsky2, Barvinsky3,Barvinsky4, Barvinsky5}. The essence of this technique consists in the possibility to convert noncovariant series in powers of gravitational potentials $h_{\mu\nu}$ into series of spacetime curvature and its derivatives with the covariant nonlocal coefficients \cite{Barvinsky1, Barvinsky2, Modesto1}.
The resulting nonlocal action generating equation $\eqref{Variational}$ begins with the quadratic order in the curvature,
\begin{equation*}
S_{NL}[g_{\mu\nu}]\,=\,-\frac{c^3}{16\pi } \int \ d^4x\ \sqrt{-g} \Big\{G^{\mu\nu}\frac{G_\Lambda^{-1}(\Box_g)}{\Box_g}R_{\mu\nu}+\mathcal{O}[R^3_{\mu\nu}]\Big\},
\end{equation*}
It can be shown that in the simplest case of constant $G(\Box_g)$ the the nonlocal action outlined above reproduces the Einstein-Hilbert action \cite {Barvinsky1, Barvinsky2}. In the context of the cosmological constant problem we aim to present in this article a precise differential coupling model which contains the degravitation properties mentioned above,
\begin{eqnarray*}
G_{\Lambda}(\Box)\,=\,\mathcal{G}_{\kappa}(\Box_g) \ \mathcal{F}_\Lambda(\Box_g),
\end{eqnarray*}
where  $\mathcal{G}_{\kappa}=\frac{G}{1-\sigma e^{\kappa\Box_g}}$ is a purely ultraviolet (UV) modification term and $\mathcal{F}_\Lambda=\frac{\Lambda \Box_g}{\Lambda \Box_g-1}$ is the nonlocal infrared (IR) contribution. We remind that $\Box_g=\nabla^{\alpha}\nabla_\alpha$ is the covariant d'Alembert operator and $G$ the Newtonian coupling constant. We see that we recover in the limit of infinitely large frequencies (vanishing wavelengths) Einstein gravity as the UV-term $\lim_{z\rightarrow +\infty} \mathcal{G}_{\kappa}(z)=G$ reduces to the Newtonian coupling constant and the IR-term $\lim_{z\rightarrow +\infty}\mathcal{F}_\Lambda=1$ goes to one. The IR-degravitation essentially comes from $\lim_{z\rightarrow 0}\mathcal{F}_\Lambda(z)=0$ while the UV-term $\lim_{z\rightarrow 0}\mathcal{G}_\kappa(z)=\frac{G}{1-\sigma}$ taken alone does not vanish in this limit. The dimensionless UV-parameter $\sigma$ is a priori not fixed, however in order to make the infrared degravitation mechanism work properly $\sigma$ should be different from one. We will see in the next chapter that we will restrain the general theory by assuming that $|\sigma|<1$ is rather small. The second UV-parameter $\kappa$ and the IR-degravitation parameter $\Lambda$ are of dimension length squared. The constant factor $\sqrt{\Lambda}$ is the cosmological scale at which the infrared degravitation process sets in. In the context of the cosmological constant problem this parameter needs to be typically of the order of the horizon size of the present visible Universe $\sqrt{\Lambda} \sim  10^{30} m$ \cite{Dvali1,Barvinsky1, Barvinsky2, Dvali2}. In addition we assume that $\sqrt{\kappa}\ll\sqrt{\Lambda}$, so that we can perform a formal series-expansion $G_\kappa(z)=\sum_{n=0}^{+\infty}\sigma^n e^{n\frac{\kappa}{\Lambda}z} $ in the UV-regime ($|z|\ll1$). The parameter $\kappa$, although named differently, was encountered in the context of various nonlocal modified theories of gravity which originate from the pursuit of constructing a UV-complete theory of quantum gravity or coming from models of noncommutative geometry \cite{Modesto1,Modesto2, Spallucci1, Sakellariadou1}. To conclude this subsection we would like to point out that in the limit of vanishing UV parameters and infinitely large IR parameter, $\lim_{\sigma,\kappa\rightarrow 0}\lim_{\Lambda\rightarrow +\infty}G_{\Lambda}(\Box_g)=G$, we recover the usual Einstein field equations.
\subsection{Degravitation of the vacuum energy:}
We intend to briefly outline the basic features of the cosmological constant problem before we return to our precise nonlocal coupling model. In the quest of generating a static universe Einstein originally introduced an additional term on the right hand side of his field equations, the famous cosmological constant. Later he dismissed this term by arguing that it was nothing else than an unnecessary complication to the field equations \cite{Einstein5, Weinberg1, Weinberg2}. However from a microscopic point of view it is not so straightforward to discard such a term, because anything that contributes to the energy density of the vacuum acts just like a cosmological constant. Indeed from a quantum point of view the vacuum is a very complex state in the sense that it is constantly permeated by fluctuating quantum fields of different origins. In agreement to Heisenberg's energy-time uncertainty principle $\Delta E \Delta t\geq 0$, one important contribution to the vacuum energy comes from the spontaneous creation of virtual particle-antiparticle pairs which annihilate shortly after \cite{Weinberg1}. Although there is some freedom in the precise computation of the vacuum energy, the most reasonable estimates range around a value of $\rho_{th}\approx 10^{111} J/m^3$ \cite{Carroll1}. Towards the end of the past century two independent research groups, the {\it High-Z Supernova Team} and the {\it Supernova Cosmology Project}, searched for distant type Ia supernovae in order to determine parameters that were supposed to provide information about the cosmological dynamics of the Universe. The two research groups were able to obtain a deeper understanding of the expansion history of the Universe by observing how the brightness of these supernovae varies with redshift. They initially expected to find signs that the expansion of the Universe is slowing down as the expansion rate is essentially determined by the energy-momentum density of the Universe. However in 1998 they published their results in two separate papers and came both independently from each other to the astonishing result that the opposite is true: the expansion of the Universe is accelerated. The supernovae results in combination with the Cosmic Microwave Background data \cite{Planck1} interpreted in terms of the Standard Model of Cosmology ($\Lambda$CDM-model) allow for a precise determination of the matter and vacuum energy density parameters of the present Universe: $\Omega_m\approx 0.3$ and $\Omega_\Lambda\approx 0.7$. 
This corresponds to an observational vacuum energy density of the order of $\rho_{ob}\sim 10^{-9} J/m^3$. Thus the supernova studies have provided direct evidence for a non zero value of the cosmological constant. These investigations together with the theoretically computed value for the vacuum energy $\rho_{th}$ lead to the famous 120-orders-of-magnitude discrepancy which makes the cosmological constant problem such a glaring embarrassment \cite{Carroll1},
\begin{equation*}
\rho_{th} \sim 10^{120} \rho_{ob}.
\end{equation*}
Most efforts in solving this problem have focused on the question why the vacuum energy is so small. However, since nobody has ever measured the energy of the vacuum by any means other than gravity, perhaps the right question to ask is why does the vacuum energy gravitates so little \cite{Dvali1,Barvinsky1,Dvali2, Barvinsky2}. In this regard our aim is not to question the theoretically computed value of the vacuum energy density, but we will rather try to see if we can find a mechanism by which the vacuum energy is effectively degravitated at cosmological scales. In order to demonstrate how the degravitation mechanism works in the context of our precise model we introduce an effective but very illustrative  macroscopic description of the vacuum energy on cosmological scales. In good agreement to cosmological observations \cite{Planck1}, we will assume that the Universe is essentially flat, so that the differential coupling operator can be approximated by its flat spacetime counterpart. We further assume that the quantum vacuum energy can be modelled, on macroscopic scales, by an almost time independent Lorentz-invariant energy process, $\langle T_{\alpha\beta}\rangle_{v}\,\simeq\, T_v \ \cos(\textbf{k}_c \cdot \textbf{x}) \ \eta_{\alpha\beta}$, where $T_v$ is the average vacuum energy density and $\textbf{k}_c=1/\mathbf{\lambda}_c$ is the three dimensional characteristic wave-vector $(|\mathbf{\lambda}_c|\gg 1)$. Moreover we suppose that the vacuum energy is homogeneously distributed throughout the whole universe so that the components of the wave-vector $k_x=k_y=k_z\,\sim 1/\lambda_c$ are the same in all three spatial directions, 
\begin{equation*}
G_\Lambda(\Box) \ \langle T_{\alpha\beta} \rangle_{v}\,
=\,\mathcal{G}(\kappa/\lambda_c^2) \ \mathcal{F}(\Lambda/\lambda_c^2) \ \langle T_{\alpha\beta} \rangle_{v},
\end{equation*}
where $\mathcal{G}(\kappa/\lambda_c^2)=\frac{G}{1-\sigma e^{-\kappa/\lambda_c^2} }$ and $\mathcal{F}(\Lambda/\lambda_c^2)=\frac{\Lambda /\lambda_c^2}{1+\ \Lambda/\lambda_c^2}$ \cite{Kragler1}. We observe that energy processes with a characteristic wavelength, much larger than the macroscopic filter scale $\lambda_c \gg \sqrt{\Lambda}$ effectively decouple from the gravitational field $\lim_{\lambda_c \rightarrow+\infty} G_\Lambda(\Box) \langle T_{\alpha\beta} \rangle_{v}\,=\, 0$. In the extreme but unlikely limit of energy processes with infinitely large frequencies, $\lim_{\lambda_c\rightarrow 0}\mathcal{G}(\kappa/\lambda^2_c)=G$, $\lim_{\lambda_c\rightarrow 0}\mathcal{F}(\Lambda/\lambda^2_c)=1$ we would recover the Newtonian coupling. 
\begin{figure}[h]
\begin{center}
\includegraphics[width=8cm,height=4.2cm]{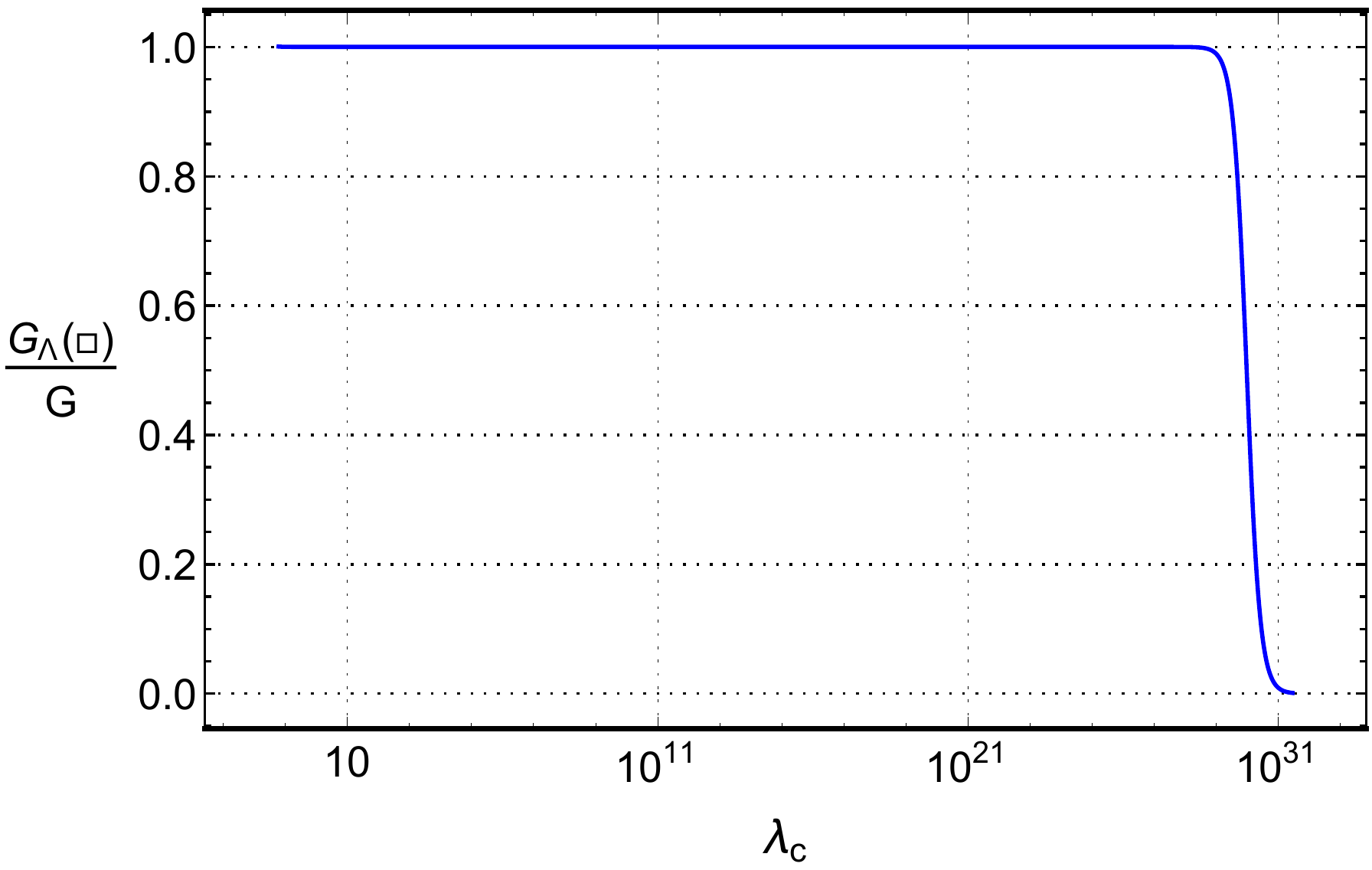} 
\end{center}
\caption{\label{Degravitation1}The function $\frac{G_\Lambda(\Box)}{G}=\frac{\mathcal{G}(\kappa/\lambda_c^2)\mathcal{F}(\Lambda/\lambda^2_c)}{G}$ is plotted against the characteristic wavelength $\lambda_c$ (m) for $\sigma=2 \ 10^{-4}$, $\kappa=5 \ 10^{-3}$ m$^2$ and $\Lambda=10^{60}$ m$^2$. A strong degravitational effect is observed for energy processes with a characteristic wavelength larger or equal to $\lambda_c=10^{29}$m.}
\end{figure}
This situation is illustrated in FIG. \ref{Degravitation1}, where we plotted the function $\big[\frac{G_\Lambda(\Box)}{G}\langle T_{\alpha\beta}\rangle_v\big][\langle T_{\alpha\beta}\rangle_v]^{-1}=\frac{\mathcal{G}(\kappa/\lambda_c^2)\mathcal{F}(\Lambda/\lambda^2_c)}{G}$ for the following UV and IR parameters, $\sigma=2\ 10^{-4}$, $\kappa=5 \ 10^{-3}$ m$^2$ and $\Lambda=10^{60}$ m$^2$, against the characteristic wavelength. We infer from FIG. \ref{Degravitation1} that in the context of our vacuum energy model we have for small characteristic wavelengths $G_\Lambda(\Box)\sim G$ while for large wavelengths of the order $\lambda_c=10^{29}$m we observe a strong degravitational effect. In the remaining chapters of this article we will investigate how much the relaxed Einstein equations are affected by the nonlocal UV-term $\mathcal{G}_\kappa(\Box_g)$. In particularly we will examine in the penultimate chapter in how far the total mass of an N-body system deviates from the purely general relativistic result. However before we embark for these computations we will shortly review the standard relaxed Einstein equations and their solutions in the context of the post-Newtonian theory. 

\section{The relaxed Einstein equations:}
The purpose of this chapter is to work out the relaxed Einstein equations and related quantities by using the very elegant Landau-Lifshitz formulation of the Einstein field equations \cite{LandauLifshitz,MisnerThroneWheeler,Will2,PatiWill1,Blanchet1,Buonanno1,Poisson,PatiWill2},
\begin{equation*}
\partial_{\mu\nu} H^{\alpha\mu\beta\nu}\,=\, \frac{16\pi G}{c^4}(-g)\ \big(T^{\alpha\beta}+t^{\alpha\beta}_{LL}\big),
\end{equation*}
where $H^{\alpha\mu\beta\nu}\,\equiv\,\mathfrak{g}^{\alpha\beta} \mathfrak{g}^{\mu\nu}-\mathfrak{g}^{\alpha\nu}\mathfrak{g}^{\beta\nu}$ is a tensor density which possesses the same symmetries as the Riemann tensor. In the Landau-Lifshitz formulation of gravity the main variables are not the components of the metric tensor $g_{\alpha\beta}$ but those of the gothic inverse metric, $\mathfrak{g}^{\alpha\beta}\,\equiv\, \sqrt{-g} \ g^{\alpha\beta}$, where $g^{\alpha\beta}$ is the inverse metric and $g$ the metric determinant \cite{LandauLifshitz, MisnerThroneWheeler,Will1,PatiWill1,PatiWill2,Blanchet1,Buonanno1,Poisson}. $T^{\alpha\beta}$ is the energy-momentum tensor of the matter source term and the Landau-Lifshitz pseudotensor,
\begin{equation*}
\begin{split}
(-g)t^{\alpha\beta}_{LL}\,=\,& \frac{c^4}{16 \pi G} \big[\partial_\lambda \mathfrak{g}^{\alpha\beta} \partial_\mu \mathfrak{g}^{\lambda\mu}-\partial_\lambda \mathfrak{g}^{\alpha\lambda}\partial_\mu \mathfrak{g}^{\beta\mu}+
\frac{1}{2} g^{\alpha\beta} g_{\lambda\mu} \partial_\rho \mathfrak{g}^{\lambda\nu}\partial_\nu \mathfrak{g}^{\mu\rho}
-g^{\alpha\lambda}g_{\mu\nu} \partial_\rho \mathfrak{g}^{\beta\nu}\partial_\lambda \mathfrak{g}^{\mu\rho}-
g^{\beta\lambda}g_{\mu\nu}\partial_\rho \mathfrak{g}^{\alpha\nu}\partial_\lambda \mathfrak{g}^{\mu\rho}\\
&\quad\quad\quad\quad\quad +g_{\lambda\mu}g^{\nu\rho}\partial_\nu \mathfrak{g}^{\alpha\lambda} \partial_\rho \mathfrak{g}^{\beta\mu}+
 \frac{1}{8} (2g^{\alpha\lambda} g^{\beta \mu}-g^{\alpha\beta}g^{\lambda\mu})(2 g_{\nu\rho} g_{\sigma\tau}-g_{\rho\sigma}g_{\nu\tau})
\partial_\lambda \mathfrak{g}^{\nu\tau}\partial_\mu \mathfrak{g}^{\rho\sigma}\big],
\end{split}
\end{equation*}
can be interpreted as an energy momentum (pseudo)tensor for the gravitational field. Although this interpretation should not be taken literally, after all it is based on a very specific formulation of the Einstein field equations, it is however supported by the fact that the $(-g)t^{\alpha\beta}_{LL}$ is quadratic in $\partial_\mu \mathfrak{g}^{\alpha\beta}$, just as the energy-momentum tensor of the electromagnetic field is quadratic in the derivative of the electromagnetic potential $\partial_\mu A^{\alpha}$. By virtue of the antisymmetry of $H^{\alpha\mu\beta\nu}$ in the last pair of indices, we have that the equation $\partial_{\beta\mu\nu} H^{\alpha\mu\beta\nu}=0$ holds as an identity. This together with the equation of the Landau-Lifshitz formulation of general relativity implies that, $\partial_\beta \big[(-g)\big(T^{\alpha\beta}+t^{\alpha\beta}_{LL}\big)\big]=0$. These are conservation equations for the total energy-momentum pseudotensor expressed in terms of the partial-derivative operator. The latter are equivalent to the energy-momentum conservation $\nabla_\beta T^{\alpha\beta}=0$ involving only the matter energy-momentum tensor and the covariant derivative operator. However there is an important conceptual difference between the two conservation relations. $\nabla_\beta T^{\alpha\beta}=0$ is a direct consequence of the local conservation of energy-momentum, as observed in a local inertial frame and is valid whether or not general relativity is the correct theory of gravity. The second conservation equation is a consequence of Einstein's field equations. If Einstein's equations are satisfied than either equation may be adopted to express energy-momentum conservation and the two statements are equivalent in this sense. It is advantageous to impose the four conditions $\partial_\beta \mathfrak{g}^{\alpha\beta}=0$ on the gothic inverse metric, known as the harmonic coordinate conditions. It is also useful to introduce the gravitational potentials defined by $h^{\alpha\beta}:=\eta^{\alpha\beta}-\mathfrak{g}^{\alpha\beta}$, where $\eta^{\alpha\beta}=diag(-,+,+,+)$ is the Minkowski metric expressed in Lorentzian coordinates \cite{Blanchet1,Blanchet3,Blanchet4,Will2, PatiWill1, PatiWill2, Buonanno1}. In terms of the potentials the harmonic coordinate conditions read $\partial_\beta h^{\alpha\beta}=0$, and in this context they are usually referred to as the harmonic gauge conditions. It is straightforward to verify that the left-hand side of the Landau-Lifshitz formulation of the Einstein field equations reduces to $\partial_{\mu\nu}H^{\alpha\mu\beta\nu}=-\Box h^{\alpha\beta}+h^{\mu\nu}\partial_{\mu\nu}h^{\alpha\beta}-\partial_\mu h^{\alpha\nu}\partial_\nu h^{\beta\mu}$, where $\Box=\eta^{\mu\nu}\partial_{\mu\nu}$ is the flat-spacetime d'Alembert operator. The right-hand side of the field equations remains essentially unchanged, but the harmonic conditions do slightly simplify the form of the Landau-Lifshitz pseudotensor, namely the first two terms in $(-g)t^{\alpha\beta}_{LL}$ vanish. Isolating the wave operator on the left-hand side and putting the remaining terms on the other side, gives rise to the formal wave equation \cite{Poisson,Will2,PatiWill1,PatiWill2,Blanchet1,Blanchet3,Blanchet4,Maggiore1,Buonanno1},
\begin{equation*}
\Box h^{\alpha\beta}\,=\,-\frac{16 \pi G}{c^4} \tau^{\alpha\beta},
\end{equation*}
where $\tau^{\alpha\beta}:=-\frac{16 \pi G}{c^4} \big[ \tau^{\alpha\beta}_m+\tau^{\alpha\beta}_{LL}+\tau^{\alpha\beta}_H\big]$ is defined as the effective energy-momentum pseudotensor composed by a matter $\tau_m^{\alpha\beta}=(-g) T^{\alpha\beta}$ contribution, the Landau-Lifshitz contribution $\tau^{\alpha\beta}_{LL}=(-g)t^{\alpha\beta}_{LL}$ and the harmonic gauge contribution, $\tau^{\alpha\beta}_H=(-g)t^{\alpha\beta}_H=\frac{c^4}{16\pi G} \big(\partial_\mu h^{\alpha\nu}\partial_\nu h^{\beta\mu}-h^{\mu\nu}\partial_{\mu\nu}h^{\alpha\beta}\big)$. It is easy to verify that because of the harmonic gauge condition this additional contribution is separately conserved, $\partial_\beta\big[(-g)t^{\alpha\beta}_H\big]=0$. This together with the conservation relation introduced previously leads to a conservation relation for the effective energy-momentum tensor $\partial_\beta \tau^{\alpha\beta}=0$. It should be noticed that so far no approximations have been introduced, so that the wave equation, together with the harmonic gauge conditions, is an exact formulation of the Einstein field equations. It is the union of these two sets of equations that is equivalent to the standard Einstein equations outlined in the previous chapter. The wave equation taken by itself, independently of the harmonic gauge condition or the conservation condition, is known as the relaxed Einstein field equation \cite{Poisson, Will2, PatiWill1, PatiWill2}. It is well known that the wave equation can be solved by the following ansatz $h^{\alpha\beta}(x)\,=\,-\frac{16G}{c^4}  \int d^4y \ G(x,y) \ \tau^{\alpha\beta}(y)$, where $\Box G(x,y)= \delta(x-y)$ is the condition for the Green function, $x=(ct,\textbf{x})$ is a field point and $y=(ct',\textbf{y})$ a source point. Inserting the retarded Green function solution $G(x,y)=\frac{-1}{4\pi}\frac{\delta(ct-ct'-|\textbf{x}-\textbf{y}|)}{|\textbf{x}-\textbf{y}|}$ into the ansatz outlined above and integrating over $y^0$ yields the formal retarded solution to the gravitational wave equation \cite{Poisson,Will2,PatiWill1,PatiWill2, Blanchet1,Blanchet3,Blanchet4,Maggiore1, Buonanno1, Maggiore2},
\begin{equation*}
h^{\alpha\beta}(t,\textbf{x})\,=\,\frac{4G}{c^4} \int d\textbf{y} \ \frac{\tau^{\alpha\beta}(y^0-|\textbf{x}-\textbf{y}|,\textbf{y})}{|\textbf{x}-\textbf{y}|},
\end{equation*}
where the domain of integration extends over the past light cone of the field point $x=(ct,\textbf{x})$. 
In order to work out this integral we need to present the important notions of near and wave zones in the general context of the wave equation and its formal solution. To do so we need to introduce the characteristic length scale of the source $r_c$ which is defined such that the matter variables vanish outside a sphere of radius $r_c$. The characteristic time scale $t_c$ is the time required for noticeable changes to occur within the source. These two important scaling quantities are related through the characteristic velocity within the source $v_c=\frac{r_c}{t_c}$. The characteristic wavelength of the radiation $\lambda_c$ produced by the source is directly related to the source's characteristic time scale $\lambda_c=c\ t_c$. This finally allows us to define the near and wave zone domains \cite{Poisson, Will2,PatiWill1,PatiWill2,Blanchet1,Maggiore1},
\begin{equation*}
\text{near-zone:} \quad r\,\ll\, \lambda_c,\quad\ \ \text{wave-zone:} \quad r\,\gg\, \lambda_c.
\end{equation*}
Thus the near zone is the region of three dimensional space in which $r=|\textbf{x}|$ is small compared with a charcateristic wavelength $\lambda_c$, while the wave zone is the region in which $r$ is large compared with this length scale. We introduce the arbitrarily selected radius $\mathcal{R}\lesssim\lambda_c$ to define the near-zone domain $\mathcal{M}:|\textbf{x}|<\mathcal{R}$. The near-zone and wave-zone domains ($\mathcal{W}:|\textbf{x}|>\mathcal{R}$) join together to form the complete light cone of some field point $y$, $\mathcal{C}(y)=\mathcal{M}(y)+\mathcal{W}(y)$. Although $\mathcal{R}$ is typically of the same order of magnitude as the characteristic wavelength of the gravitational radiation, it was shown in \cite{Poisson,Will2,PatiWill1,PatiWill2} that the precise choice of $\mathcal{R}$ is irrelevant because we observe a mutual cancellation between terms being proportional to $\mathcal{R}$ coming from the near and wave zones. While the gravitational potentials originating from the two different intgration domains will individually depend on the cutoff radius their sum is guaranteed to be $\mathcal{R}$-independent and we will therefore discard such terms in the remaining part of this article \cite{Poisson,Will2,PatiWill1}. The gravitational potentials behave very differently in the two zones: in the near zone the difference between the retarded time $\tau=t-r/c$ and $t$ is small, so that the field retardation is unimportant. In the wave zone the difference between $\tau$ and $t$ is large and time derivatives are comparable to spatial derivatives. The post-Minkowskian theory is an approximation method that will not only reproduce the predictions of Newtonian theory but is a method that can be pushed systematically to higher and higher order to produce an increasingly accurate description of a weak gravitational field $||h^{\alpha\beta}||<1$. In this sense the metric of the spacetime will be constructed by considering a formal expansion of the form $h^{\alpha\beta}=Gk_1^{\alpha\beta}+G^2k_2^{\alpha\beta}+G^3k_3^{\alpha\beta}+...$ for the gravitational potentials. Such an approximation in powers of $G$ is known as post-Minkowskian expansion with the aim to obtain, at least in a useful portion of spacetime, an acceptable approximation to the true metric \cite{Poisson}. The spacetime deviates only moderately from Minkowski spacetime and we can construct the spacetime metric $g_{\alpha\beta}$ from the gravitational potentials,
\begin{equation*}
g_{\alpha\beta}=\eta_{\alpha\beta}+h_{\alpha\beta}-\frac{1}{2}h\eta_{\alpha\beta}+h_{\alpha\mu}h^{\mu}_\beta-\frac{1}{2}hh_{\alpha\beta}+\Big(\frac{1}{8}h^2-\frac{1}{4}h^{\mu\nu}h_{\mu\nu}\Big)\eta_{\alpha\beta}+\mathcal{O}(G^3),
\end{equation*}
where the indices on $h^{\alpha\beta}$ are lowered with the Minkowski metric $h_{\alpha\beta}=\eta_{\alpha\mu}\eta_{\beta\nu}h^{\mu\nu}$ and $h=\eta_{\mu\nu}h^{\mu\nu}$. The method is actually so successful that it can handle fields that are not so weak at all and therefore be employed for a description of gravity at a safe distance from neutron stars or even binary-black hole systems. The link between the spacetime metric $g_{\alpha\beta}$ and the gravitational potentials is provided by the gothic inverse metric $\mathfrak{g}^{\alpha\beta}=\eta^{\alpha\beta}-h^{\alpha\beta}$ \cite{Poisson, Will2, PatiWill1, PatiWill2,Blanchet1,Maggiore1, Buonanno1} and the metric determinant is given by $(-g)=1-h+\frac{1}{2}h^2-\frac{1}{2}h^{\mu\nu}h_{\mu\nu}+\mathcal{O}(G^3)$. The post-Minkowskian expansion of the metric, adjusted to the context of our modified theory of gravity, will be frequently used in the next chapters. In what follows we will assume that the matter distribution of the source is deeply situated within the near zone $r_c\ll \lambda_c$, where we remind that $r_c$ is the characteristic length scale of the source. It is straightforward to observe that this equation is tantamount to a slow motion condition $v_c\ll c$ for the matter source term. The post-Newtonian theory (pN) is an approximation method to the theory of general relativity that incorporates both weak-field and slow-motion. The dimensionless expansion parameter in this approximation procedure is $(Gm_c)/(c^2r_c)=v^2_c/c^2$, where $m_c$ is the characteristic mass of the system under consideration. In the context of this article, we are primarily interested in the near-zone piece of the gravitational potentials $h^{ab}_{\mathcal{N}}$. It can be shown \cite{Poisson, Will2, PatiWill1, PatiWill2} that the formal near-zone solution to the wave equation, for a far-away wave-zone field point ($|\textbf{x}|\gg\lambda_c$) can be rephrased in the following way,
\begin{equation*}
h^{ab}_\mathcal{N}=\frac{4G}{c^4r}\sum_{l=0}^{+\infty}\frac{n_L}{l!}\Big(\frac{d}{du}\Big)^l\int_{\mathcal{M}} d\textbf{y} \ \tau^{ab}(u,\textbf{y}) \ y^L+\mathcal{O}(r^{-2}),
\end{equation*}
by expanding the ratio $\frac{\tau^{\alpha\beta}(t-|\textbf{x}-\textbf{y}|/c,\textbf{y})}{|\textbf{x}-\textbf{y}|}=\frac{1}{r} \ \sum_{l=0}^\infty \frac{y^L}{l!}  \ n_L \ \Big(\frac{\partial}{\partial u}\Big)^l \ \tau^{\alpha\beta}(u,\textbf{y})+\mathcal{O}(1/r^2)$ in terms of the retarded time $u=c\tau$ and the unit radial vectors $\textbf{n}=\frac{\textbf{x}}{r}$. The far away wave zone is characterized by the fact that only leading order terms $1/r$ need to be retained and $y^Ln_L=y^{j1}\cdots y^{jl}n_{j1}\cdots n_{jl}$. We will return to this expansion in chapter three where we outline a similar computation in the framework of the nonlocally modified theory of gravity presented in the introduction of the present article. We model the material source term by a collection of N-fluid balls with negligible pressure, $T^{\alpha \beta}=\rho\ u^\alpha u^\beta$, where $\rho\big(m_A,\textbf{r}_A(t)\big)$ is the energy-density and $u^\alpha=\gamma_A (c,\textbf{v}_A)$ is the relativistic four-velocity of the fluid ball with mass $m_A$ and individual trajectory $r_A(t)$. Further details on this important quantity can be withdrawn from the appendix-section related to this chapter. The slow-motion condition gives rise to a hierarchy between the components of the energy-momentum tensor $T^{0b}/T^{00}\sim v_c/c$ and $T^{ab}/T^{00}\sim (v_c/c)^2$, where we used the approximate relations $T^{00}\approx \rho\ c^2$, $T^{0b}\approx\rho\ v^b c$, $T^{ab}\approx \rho\ v^a v^b$ and $\textbf{v}$ is the three-dimensional velocity vector of the fluid balls. A glance at the relaxed Einstein equations reveals that this hierarchy is inherited by the gravitational potentials $h^{0b}/h^{00}\sim v_c/c$, $h^{ab}/h^{00}\sim (v_c/c)^2$. Taking into account the factor $c^{-4}$ in the field equations, we have for the potentials $h^{00}=\mathcal{O}(c^{-2})$, $h^{0b}=\mathcal{O}(c^{-3})$ and $h^{ab}=\mathcal{O}(c^{-4})$, where $c^{-2}$ is a post-Newtonian expansion parameter. We remind that this notation serves only as a powerful mnemonic to judge the importance of various terms inside a post-Newtonian expansion, while the real dimensionless expansion parameter is rather $(Gm_c)/(c^2r_c)=v^2_c/c^2$. The precise shape of the 1.5 post-Newtonian time-time matter component of the energy-momentum pseudotensor, 
\begin{equation*}c^{-2}(-g)T^{00}=\sum_A m_A\ \Big[1+\frac{1}{c^2}(\frac{\textbf{v}^2}{2}+3U)\Big]\ \delta(\textbf{x}-\textbf{r}_A)+\mathcal{O}(c^{-4}),
\end{equation*}
is worked out in the appendix-section related to this chapter. $U$ is the Newtonian potential of a N-body system with point masses $m_A$ and $h^{00}=\frac{4}{c^2}U+\mathcal{O}(c^{-4})$ is the corresponding gravitational potential at the 1.5 post-Newtonian order of accuracy. Another important relation, that will be frequently used in chapter five, is the time-time component of the Landau-Lifshitz tensor $\tau^{00}_{LL}=(-g)t^{00}_{LL}$ worked out to the required degree of accuracy \cite{Poisson,Will2,PatiWill1,PatiWill2}. Here again we will see that in the context of our modified theory of gravity, we need to adapt the result,
\begin{equation*}
c^{-2}(-g)t^{00}_{LL}\,=\,-\frac{7}{8\pi G c^2}\ \big[\partial_pU\partial^pU\big]+\mathcal{O}(c^{-4}).
\end{equation*}
Further computational details regarding the derivation of this quantity can be inferred from the appendix-section related to this chapter. Using the information gathered previously we see that the harmonic gauge contribution is beyond the 1.5 post-Newtonian order of accuracy $c^{-2}\tau_H^{00}=\mathcal{O}(c^{-4})$. To conclude this chapter we aim to introduce the total mass $M_V=c^{-2}\int_Vd\textbf{x}\ (-g)(T^{00}+t_{LL}^{00})$ contained in a three-dimensional region $V$ and bounded by the surface $S$. The latter is a direct consequence of the energy-momentum conservation and we will return to this integral relation in chapter five.
To conclude this chapter we would like to mention that the approach which we use to integrate the wave equation is usually referred to as the Direct Integration of the Relaxed Einstein equations or DIRE approach for short. An alternative method, based on a formal multipolar expansion of the potential outside the source was nicely outlined in \cite{Blanchet1,Blanchet5, BlanchetDamour1}. Additional information on these and related issues together with applications to binary-systems can be found in a vast number of excellent articles\cite{BlanchetDamourIyerWillWiseman,DamourJaranowskiSchaefer1,BlanchetDamourIyer,BlanchetDamourEsposito-FareseIyer1,BlanchetDamourEsposito-FareseIyer2,DamourJaranowskiSchaefer2}.

\section{The modified relaxed Einstein equations:}
The main objective of this section is to work out and to solve the nonlocally modified wave equation. The latter merely arises from the quest of rewriting the relaxed Einstein equation, containing the effective energy-momentum tensor $\mathcal{T}^{\alpha\beta}= \frac{G_\Lambda(\Box)}{G} \ T^{\alpha\beta}$, in such a way that it can be solved most easily. This goal can be achieved by spreading out some of the differential complexity inside the effective energy-momentum tensor to both sides of the differential equation. We will see that the distribution of nonlocality between both sides of the wave equation will be done in a way that the gravitational potentials $h^{\alpha\beta}$ can be evaluated similarly to the purely general relativistic case. However before we can come to the actual derivation of the modified wave equation we first need to carefully prepare the grounds by setting in place a couple of important preliminary results.
\subsection{The effective energy-momentum tensor:}
The major difference between our nonlocally modified theory and the standard theory gravity lies in the way in which the energy (matter or field energy) couples to the gravitational field. In the purely Einsteinian theory the (time-dependent) distribution of energy is translated via the constant coupling $G$ into spacetime curvature. We saw in the introduction that in the case of the modified theory the coupling-strength itself varies according to the characteristic wavelength $\lambda_c$ of the source term under consideration. From a strictly formal point of view however, the cosmologically modified field equations can be formulated in a very similar way to Einstein's field equations,
\begin{equation*}
G^{\alpha\beta}\,=\, \frac{8\pi}{ c^4}\ G \ \mathcal{T}^{\alpha\beta}.
\end{equation*}
$G^{\alpha\beta}$ is the usual Einstein tensor and $\mathcal{T}^{\alpha\beta}$ is the modified energy-momentum tensor outlined in the introduction of this chapter. We see that this formulation is possible only because the nonlocal modification can be put entirely into the source term $\mathcal{T}^{\alpha\beta}$, leaving in this way the geometry ($G^{\alpha\beta}$) unaffected. In this regard we can easily see that, by virtue of the contracted Bianchi identities $\nabla_\beta G^{\alpha\beta}=0$, the modified energy-momentum tensor is conserved $\nabla_\beta \mathcal{T}^{\alpha\beta}=0$. This allows us to use the Landau-Lifshitz formalism introduced previously by simply replacing the energy-momentum tensor $T^{\alpha\beta}$, inside the relaxed Einstein field equations, through its nonlocal counterpart, 
\begin{equation*}
\Box h^{\alpha\beta}\,=\,-\frac{16\pi G}{c^4} \ (-g) \ \Big[\mathcal{T}^{\alpha\beta}+t^{\alpha\beta}_{LL}+t^{\alpha\beta}_H\Big].
\end{equation*}
Instead of trying to integrate out by brute force the nonlocally modified relaxed Einstein field equations, we rather intend to bring part of the differential complexity, stored inside the effective energy momentum tensor, to the left-hand-side of the field equation. These efforts will finally bring us to an equation that will be more convenient to solve. Loosely speaking we aim to separate inside the nonlocal covariant differential coupling the flat spacetime contribution from the the curved one. In this way we can rephrase the relaxed Einstein equations in a form that we will eventually call the nonlocally modified wave equation or effective relaxed Einstein equation. This new equation will have the advantage that the nonlocal complexity will be distributed to both sides of the equation and hence it will be easier to work out its solution to the desired post-Newtonian order of accuracy. In this context we aim to rewrite the covariant d'Alembert operator $\Box_g$ in terms of a flat spacetime contribution $\Box$ plus an additional piece $w$ depending on the gravitational potentials $h^{\alpha\beta}$. The starting point for the splitting of the differential operator $\Box_g=\nabla_\alpha\nabla^\alpha$ is the well known relation \cite{Poisson,Woodard1, Weinberg1, Maggiore2},
\begin{equation*}
\Box_g\,=\,\frac{1}{\sqrt{-g}}\partial_\mu\big(\sqrt{-g}g^{\mu\nu}\partial_\nu)\,=\,\Box+w(h,\partial),
\end{equation*}
where $\Box=\partial^\alpha\partial_\alpha$ is the flat spacetime d'Alembert operator and the differential operator function $w(h,\partial)\,=\,-h^{\mu\nu}\partial_\mu\partial_\nu+\tilde{w}(h)\Box-\tilde{w}(h)h^{\mu\nu}\partial_{\mu}\partial_\nu+\mathcal{O}(G^4)$ is composed by the four-dimensional spacetime derivatives $\partial_\beta$ and the potential function $\tilde{w}(h)= \frac{h}{2}-\frac{h^2}{8}+\frac{h^{\rho\sigma}h_{\rho\sigma}}{4}+\mathcal{O}(G^3)$. We remind that the actual expansion parameter in a typical situation involving a characteristic mass $m_c$ confined to a region of characteristic size $r_c$ is the dimensionless quantity $Gm_c/(c^2 r_c)$. The result above was derived by employing the post-Minkoskian expansion of the metric $g_{\alpha\beta}$ in terms of the gravitational potentials \cite{Poisson, Will2, PatiWill1, PatiWill2,Blanchet1} outlined in the previous chapter. Further computational details can be found in the appendix relative to this chapter. With this result at hand we are ready to split the nonlocal gravitational coupling operator $G(\Box_g)$ into a flat spacetime contribution $G(\Box)$ multiplied by a piece $\mathcal{H}(\Box,w)$ that may contain correction terms originating from a possible curvature of spacetime,
\begin{equation*}
\mathcal{T}^{\alpha\beta}\,=\,G(\Box) \ \mathcal{H}(\Box,w) \ T^{\alpha\beta},
\end{equation*}
For astrophysical processes confined to a rather small volume of space $r_c\ll \sqrt{\Lambda}$ we can reduce the nonlocal coupling operator $G_{\Lambda}(\Box_g)$ to its ultraviolet component $G(\Box_g)=$G$ \big[1-\sigma e^{\kappa\Box_g}\big]^{-1}$ only. Using the relation for the general covariant d'Alembert operator, we can split the differential UV-coupling into two separate contributions, 
\begin{equation*}
G(\Box)\,=\,\frac{1}{1-\sigma e^{\kappa\Box}},\quad \ \mathcal{H}(w,\Box)\,=\,1+\sigma \frac{e^{\kappa\Box}}{1-\sigma e^{\kappa\Box}} \sum_{n=1}^{+\infty} \frac{\kappa^n}{n!} w^n+...
\end{equation*}
The price to pay to obtain such a concise result is to assume that the modulus of the dimensionless parameter $\sigma$ has to be smaller than one ($|\sigma|<1$). Here again the reader interested in the computational details is referred to the appendix where a detailed derivation of this result can be found. It will turn out that the splitting of the nonlocal coupling operator, into two independent pieces, will be of serious use when it comes to the integration of the relaxed Einstein equations. For later purposes we need to introduce the effective curvature energy-momentum tensor,
\begin{equation*}
\mathcal{B}^{\alpha\beta}\,=\,\mathcal{H}(\Box,w) \ T^{\alpha\beta}.
\end{equation*}
It is understood that a nonlocal theory involves infinitely many terms. However in the context of a post-Newtonian expansion, the newly introduced curvature energy-momentum tensor $\mathcal{B}^{\alpha\beta}$, can be truncated at a certain order of accuracy. In this sense the first four leading terms (appendix) of the effective curvature energy-momentum tensor are,

\begin{minipage}{0.6\textwidth}
\begin{equation*}
\begin{split}
\mathcal{B}^{\alpha\beta}_1\,&=\,\Big[\frac{\tau^{\alpha\beta}_m}{(-g)}\Big],\\
\mathcal{B}^{\alpha\beta}_2\,&=\,\epsilon e^{\kappa\Box}\Big[\frac{w}{1-\sigma e^{\kappa \Box}}\Big] \Big[\frac{\tau^{\alpha\beta}_m}{(-g)}\Big],
\end{split}
\end{equation*}
\end{minipage}
\hspace{-2.9cm}
\begin{minipage}{0.4\textwidth}
\begin{equation*}
\begin{split}
\mathcal{B}^{\alpha\beta}_3\,&=\,\epsilon\frac{\kappa}{2} e^{\kappa\Box}\Big[\frac{w^2}{1-\sigma e^{\kappa \Box}}\Big] \Big[\frac{\tau^{\alpha\beta}_m}{(-g)}\Big],\\
\mathcal{B}^{\alpha\beta}_4\,&=\,\epsilon\frac{\kappa^2}{3!} e^{\kappa\Box}\Big[\frac{w^3}{1-\sigma e^{\kappa \Box}}\Big] \Big[\frac{\tau^{\alpha\beta}_m}{(-g)}\Big].
\end{split}
\end{equation*}
\end{minipage}
$\newline$

For clarity reasons we introduced the parameter $\epsilon=\kappa \sigma$ of dimension length squared. Moreover we will see in the next chapter that the infinitely many remaining terms are in the sense of a post-Newtonian expansion beyond the degree of accuracy at which we aim to work at in this article. To conclude this subsection we would like to point out that the leading term in the curvature energy-momentum tensor can be reduced to the matter source term, $\mathcal{B}^{\alpha\beta}_1=T^{\alpha\beta}$.  
\subsection{The nonlocally modified wave equation:}
We are now ready to come to the main part of this chapter in which we intend to work out the nonlocally modified wave equation. As already hinted in the introduction of this chapter, the modified wave equation naturally originates from the quest of sharing out some of the complexity of the nonlocal coupling operator $G(\Box_g)$ to both sides of the relaxed Einstein equations. We have shown in the previous subsection (and in the corresponding appendix-section) that it is possible to split the nonlocal coupling operator, acting on the matter source term $T^{\alpha\beta}$, into a flat space contribution $G(\Box)$ multiplied by a highly nonlinear differential piece $\mathcal{H}(\Box,w)$. We aim to summarize first what this means for the effective energy-momentum tensor, $\mathcal{T}^{\alpha\beta}=G(\Box) \ \mathcal{H}(w,\partial) \ T^{\alpha\beta}$. In the pursuit of removing some of the differential complexity from the effective energy-momentum tensor $\mathcal{T}^{\alpha\beta}$ we will apply the inverse flat spacetime operator $G^{-1}(\Box)$ to both sides of the relaxed Einstein field equation, $G^{-1}(\Box) \ \Box h^{\alpha\beta}\,=\,-\frac{16 \pi G}{c^4} \ G^{-1}(\Box) \big[(-g)\mathcal{T}^{\alpha\beta}+\tau_{LL}^{\alpha\beta}+\tau_H^{\alpha\beta}\big]$. We will see that it is precisely this mathematical operation which will finally lead us to the modified wave equation,
\begin{eqnarray*}
\Box_{c} \ h^{\alpha\beta}(x)\,=\, -\frac{16 \pi G}{c^4}N^{\alpha\beta}(x).
\end{eqnarray*}
where $\Box_{c}$ is the effective d'Alembert operator $\Box_{c}=\big[1-\sigma e^{\kappa\Delta}\big] \ \Box$. $N^{\alpha\beta}$ is a pseudotensorial quantity  which we will call in the remaining part of this article the effective energy-momentum pseudotensor, $N^{\alpha\beta}=G^{-1}(\Box) \big[(-g)\mathcal{T}^{\alpha\beta}+\tau_{LL}^{\alpha\beta}+\tilde{\tau}_H^{\alpha\beta}\big]$, where $\tilde{\tau}^{\alpha\beta}_m=(-g)\mathcal{T}^{\alpha\beta}$ is the effective matter pseudotensor, $\tau_{LL}^{\alpha\beta}=(-g)t_{LL}^{\alpha\beta}$ is the Landau-Lifshitz pseudotensor and $\tilde{\tau}_H^{\alpha\beta}=(-g)t^{\alpha\beta}_H+G(\Box)\mathcal{O}^{\alpha\beta}(h)$ is the effective harmonic gauge pseudotensor where $\mathcal{O}^{\alpha\beta}(h)=-\sigma\sum_{n=1}^{+\infty}\frac{(\kappa)^n}{n!}\partial^{2n}_0 e^{\kappa\Delta} \Box h^{\alpha\beta}$ is the iterative post-Newtonian potential correction contribution. This term is added to the right-hand-side of the wave equation very much like the harmonic gauge contribution is added to the right-hand-side for the standard relaxed Einstein equation \cite{Poisson, Will2, PatiWill1, PatiWill2, Blanchet1}. It should be noticed that the modified d'Alembert operator $\Box_{c}$ is of the same post-Newtonian order than the standard d'Alembert operator, $\Box_c=\mathcal{O}(c^{-2})$ and reduces to the usual one in the limit of vanishing UV modification parameters $\lim_{\sigma,\kappa \rightarrow 0}\ \Box_{c}\,=\, \Box$. In the same limits the effective pseudotensor $N^{\alpha\beta}$ reduces to the general relativistic one, $\lim_{\sigma,\kappa \rightarrow 0}  \ N^{\alpha\beta}=\tau^{\alpha\beta}$. The second limit is less straight forward, but from the precise form of $\mathcal{T}^{\alpha\beta}$ as well as from the inverse differential operator $G^{-1}(\Box)$ we can see that we recover usual effective energy-momentum pseudotensor $\tau^{\alpha\beta}=\tau^{\alpha\beta}_m+\tau^{\alpha\beta}_{LL}+\tau^{\alpha\beta}_H$. Further conceptual and computational details on this very important quantities will be provided in the next chapter. At the level of the wave equations, these two properties can be summarized by the following relation,
\begin{equation*}
\Box_c \ h^{\alpha\beta}(x)\,=\, -\frac{16 \pi G}{c^4}N^{\alpha\beta}(x) \ \ \underset{\sigma,\kappa\rightarrow 0}{\Longrightarrow} \ \ \Box \ h^{\alpha\beta}(x)\,=\, -\frac{16 \pi G}{c^4}\tau^{\alpha\beta}(x).
\end{equation*}
In order to solve this equation we will use, in analogy to the standard wave equation, the following ansatz, $h^{\alpha\beta}(x)\,=\,-\frac{16 \pi G_N}{c^4} \int d^4y \ G(x-y) \ N^{\alpha\beta}(y)$ together with the identity for the effective Green function, $\Box_{c} G(x-y)\,=\,\delta(x-y)$, to solve for the potentials $h^{\alpha\beta}$ of the modified wave equation. Following the usual procedure \cite{Poisson,Maggiore1,Buonanno1} we obtain the Green function in momentum space,
\begin{equation*}
G(k)\,=\, \frac{1}{(k^0)^2-|\textbf{k}|^2}+\sigma \ \frac{ \ e^{-\kappa|\textbf{k}|^2}}{(k^0)^2-|\textbf{k}|^2}+\cdots.
\end{equation*}
In the remaining part of this article we will retain only the first two leading terms. It should be noticed that the first of these two contributions will eventually give rise to the usual Green function. Additional terms could have been added but as the dimensionless parameter $\sigma<1$ is by assumption strictly smaller than one the remaining terms, each by itself, contribute less than those that we have retained. Further computational details can be found in the appendix related to this chapter. These considerations finally permit us to work out an expression for the retarded Green function,
\begin{equation*}
G_r(x-y)\,=\,G_r^{GR}+G_r^{NL},
\end{equation*}
where $G_r^{GR}=\frac{-1}{4\pi}\frac{\delta(x^0-|\textbf{x}-\textbf{y}|-y^0)}{|\textbf{x}-\textbf{y}|}$ is the well known retarded Green function and
$G_r^{NL}=\frac{-1}{4\pi}\frac{1}{|\textbf{x}-\textbf{y}|}\frac{\sigma}{2\sqrt{\kappa\pi}}e^{-\frac{(x^0-|\textbf{x}-\textbf{y}|-y^0)^2}{4\kappa}}$ is the nonlocal correction term. In this way we are able to recover in the limit of vanishing modification parameters the usual retarded Green function, $\lim_{\sigma,\kappa \rightarrow 0} \ G_r(x-y)=G_r^{GR}$. In addition it should be pointed out that we have, by virtue of the exponential representation of the dirac distribution, $\lim_{\kappa\rightarrow 0} \frac{1}{2\sqrt{\kappa\pi}}e^{-\frac{(x^0-|\textbf{x}-\textbf{y}|-y^0)}{4\kappa}}\,=\, \delta(x^0-|\textbf{x}-\textbf{y}|-y^0)$. In analogy to the purely general relativistic case, we can write down the formal solution to the modfied wave equation,
\begin{equation*}
h^{\alpha\beta}(x)\,=\,  \frac{4 \ G}{c^4}  \int d\textbf{y} \ \frac{N^{\alpha\beta}(x^0-|\textbf{x}-\textbf{y}|,\textbf{y})}{|\textbf{x}-\textbf{y}|}.
\end{equation*}
The retarded effective pseudotensor can be decomposed into two independent pieces according to the two contributions coming from the retarded Green function, $N^{\alpha\beta}(x^0-|\textbf{x}-\textbf{y}|,\textbf{y})=\mathcal{D} N^{\alpha\beta}(y^0,\textbf{y})+\sigma \mathcal{E} N^{\alpha\beta}(y^0,\textbf{y})$, where for later convenience we introduced the following two integral operators, $\mathcal{D}= \int dy^0 \ \delta(x^0-|\textbf{x}-\textbf{y}|-y^0)$ and $\mathcal{E}=\int dy^o \ \frac{1}{2\sqrt{\pi \kappa}} e^{-\frac{(x^o-|\textbf{x}-\textbf{y}|-y^o)^2}{4\kappa}}$. We would like to conclude this subsection by taking a look at the modified Newtonian potential which is frequently used in post-Newtonian developments. The corresponding gravitational potential, $h^{00}(x)=\frac{4}{c^2} V(x)$, is obtained from the integral outlined above where the leading order contribution of the effective energy-momentum pseudotensor was used $N^{00}=\sum_A m_A  c^2 \ \delta(\textbf{x}-\textbf{r}_A)+\mathcal{O}(c^{-1})$. From this we obtain the modified Newtonian potential for a N-body-system $V(x)=\sum_A  \ \frac{G\tilde{m}_A}{|\textbf{x}-\textbf{r}_A|}=(1+\sigma) \ U(\textbf{x})$, where $U(x)$ is the standard Newtonian potential term and and $\tilde{m}_A=(1+\sigma) \ m_A$ is the effective mass of the body $A$. Further computational details are provided in the appendix related to this chapter. It should be noticed that the usual Newtonian potential is recovered in the limit of vanishing $\sigma$. Experimental results \cite{Chiaverini1, Kapner1} from deviation measurements of the Newtonian law at small length scales ($\sim 25 \mu m$) suggest that the dimensionless correction constant needs to be of the order $\sigma \lesssim 10^{-4}$. We see that this experimental bound confirms our theoretical assumption of a small dimensionless parameter $\sigma$.
\subsection{Solution for a far away wave-zone field point:}
In the context of astrophysical systems \cite{Wex1, LIGO1} we can restrain the general solution for the gravitational potentials to a situation in which the potentials will be evaluated for a far away wave-zone field point ($|\mathbf{x}|\gg \lambda_c$). Furthermore we will only focus in this article on the near zone energy-momentum contribution to the gravitational potentials $h^{ab}_\mathcal{N}$. In order to determine the precise form of the spatial components of the near-zone gravitational potentials we need to expand the ratio inside the formal solution \cite{Poisson, Will2, PatiWill1} in terms of a power series,
\begin{equation*}
\begin{split}
 \frac{N^{ab}(x^0-|\textbf{x}-\textbf{y}|,\textbf{y})}{|\textbf{x}-\textbf{y}|}\,
%&=\, \sum_{l=0}^{\infty} \frac{(-1)^l}{l!} \textbf{y}^L\partial_L \Big[\frac{N^{\alpha\beta}(x^0-r,\textbf{y})}{r}\Big]\\
=\,\frac{1}{r} \ \sum_{l=0}^\infty \frac{y^L}{l!}  \ n_L \ \Big(\frac{\partial}{\partial u}\Big)^l \ N^{ab}(u,\textbf{y})+\mathcal{O}(1/r^2),
\end{split}
\end{equation*}
where $u=c\tau$ and $\tau=t-r/c$ is the retarded time. The distance from the matter source term's center of mass to the far away field point is given by $r=|\textbf{x}|$ and its derivative with respect to spatial coordinates is $\frac{\partial r}{\partial x^a}=n^a$, where $n^a=\frac{x^a}{r}$ is the a-th component of the unit radial vector. The far away wave zone is characterized by the fact that only leading order terms $1/r$ need to be retained and $y^Ln_L=y^{j1}\cdots y^{jl}n_{j1}\cdots n_{jl}$. More technical details can be found in the appendix relative to this subsection. By introducing the far away wave zone expansion of the effective energy-momentum-distance ratio into the formal solution of the potentials we finally obtain the near zone contribution to the gravitational potentials for a far away wave zone field point in terms of the retarded derivatives,
\begin{equation*}
\begin{split}
h^{ab}_{\mathcal{N}}(x)\,&=\, \frac{4 G}{c^4 r} \sum_{l=0}^\infty \frac{n_L}{l!} \Big(\frac{\partial}{\partial u}\Big)^l \Big[ \int_{\mathcal{M}} d\textbf{y} \ N^{ab}(u,\textbf{y}) \ y^L\Big]+\mathcal{O}(r^{-2}),
%&=\,\frac{4 G}{c^4 r}\Big[\int_{\mathcal{M}} d\textbf{y} \ N^{ab}(u,\textbf{y}) + \frac{n_c }{c} \frac{\partial}{\partial t_r} \int_{\mathcal{M}} d\textbf{y} \ N^{ab}(u,\textbf{y}) \ y^c\\
%&\quad+\frac{n_c n_d}{2c^2} \frac{\partial^2}{\partial t_r^2} \int_{\mathcal{M}} d\textbf{y} \ N^{ab}(u,\textbf{y})y^c \ y^d\\
%&\quad +\frac{n_c n_d n_e}{6c^3} \frac{\partial^3}{\partial t^3_r} \int_{\mathcal{M}} d\textbf{y} \ N^{ab}(u,\textbf{y}) y^c \ y^d \ y^e+[l\ge4]\Big]+\mathcal{O}(r^{-2})
\end{split}
\end{equation*}
where $\mathcal{M}$ is the three-dimensional near zone integration domain (sphere) defined by $|\textbf{x}| <\mathcal{R}\leq \lambda_c$. Further computational details can be inferred from the related appendix-subsection. In order to unfold the near zone potentials in terms of the radiative multipole moments we need to introduce the modified conservation relations. They originate, like for the purely general relativistic case \cite{Poisson, Will2}, from the conservation of the effective energy momentum pseudotensor, $\partial_\beta N^{\alpha\beta}=0$. This quantity is indeed conserved because we can store the complete differential operator complexity inside the effective energy-momentum tensor $\mathcal{T}^{\alpha\beta}=G(\Box_g)T^{\alpha\beta}$. We saw in the previous chapter that as long as the the geometry is not affected by the modification ($\nabla_\beta G^{\alpha\beta}=0$) we have, no matter what the precise form of the energy-momentum tensor is, the following conservation relation, $\partial_\beta N^{\alpha\beta}=G^{-1}(\Box) \ \partial_\beta \big[(-g)\mathcal{T}^{\alpha\beta}+\tau^{\alpha\beta}_{LL}+\tilde{\tau}^{\alpha\beta}_H\big]=0$. It should be noticed that similarly to the harmonic gauge contribution $\partial_\beta t_H^{\alpha\beta}=0$ the iterative potential contribution is separately conserved $\partial_\beta \mathcal{O}^{\alpha\beta}(h)=0$ because of the harmonic gauge condition. As the linear differential operator with constant coefficients $G^{-1}(\Box)$, commutes with the partial derivative ($[G(\Box)^{-1},\partial_\beta]=0$), we can immediately conclude for the conservation of the effective energy-momentum pseudotensor and deduce the modified conservation relations, 
\begin{equation*}
\begin{split}
N^{ab}\,&=\, \frac{1}{2} \frac{\partial^2}{\partial u^2} (N^{00} x^a x^b)+\frac{1}{2} \partial_c (N^{ac} x^b+N^{bc} x^a-\partial_d N^{cd} x^a x^b),\\
N^{ab} x^c\,&=\, \frac{1}{2} \frac{\partial}{\partial u} (N^{0a} x^b x^c+N^{0b} x^a x^c -N^{0c} x^a x^b)+\frac{1}{2} \partial_d(N^{ad} x^b x^c +N^{bd} x^a x^c-N^{cd} x^a x^b).
\end{split}
\end{equation*}
A more detailed derivation of the latter is provided in the appendix related to this chapter. Finally we can rephrase the spatial components of the near-zone gravitational potentials for a far away wave zone field point in terms of the radiative multipole moments,
\begin{equation*}
h^{ab}_{\mathcal{N}}\,=\, \frac{2G}{c^4 r} \frac{\partial^2}{\partial \tau^2} \Big[Q^{ab}+Q^{abc} \ n_c+Q^{abcd} \ n_c n_d+\frac{1}{3}Q^{abcde} \ n_c n_d n_e+[l\geq 4]\Big]+\frac{2 G}{c^4 r}\left[P^{ab}+P^{abc} n_c\right]+\mathcal{O}(r^{-2}).
\end{equation*}
We see that in analogy to the purely general relativistic case \cite{Poisson, Maggiore1, Will2, PatiWill1, PatiWill2, Buonanno1, Blanchet1} the leading order term is proportional to the second derivative in $\tau$ of the radiative quadrupole moment . The first four modified radiative multipole moments are, 
\begin{minipage}{0.7\textwidth}
\begin{equation*}
\begin{split}
Q^{ab}\,=&\, \frac{1}{c^2} \int_{\mathcal{M}} N^{00} y^a y^b d\textbf{y},\\
Q^{abc}\,=&\, \frac{1}{c^2} \int_{\mathcal{M}} (N^{0a} y^b y^c+N^{0b} y^a y^c- N^{0c} y^a y^b ) \ d\textbf{y},
\end{split}
\end{equation*}
\end{minipage}
\hspace{-2.3cm}
\begin{minipage}{0.3\textwidth}
\begin{equation*}
\begin{split}
Q^{abcd}\,=&\, \frac{1}{c^2} \int_{\mathcal{M}} N^{ab} y^c y^d d\textbf{y},\\
Q^{abcde}\,=&\, \frac{1}{c^2} \int_{\mathcal{M}} N^{ab} y^c y^d y^e d\textbf{y}.
\end{split}
\end{equation*}
\end{minipage}

It can be shown that the surface terms $P^{ab}$ and $P^{abc}$, outlined in the appendix, will give rise to $\mathcal{R}$-dependent contributions only. These terms will eventually cancel out with contributions coming from the wave-zone as was shown in \cite{Will2}.
\section{The effective energy-momentum pseudotensor:}
In the previous chapter we transformed the original wave equation, in which all the nonlocal complexity was stored inside the effective energy-momentum tensor $\mathcal{T}^{\alpha\beta}$, into a modified wave equation which is much easier to solve. This effort gave rise to a new pseudotensorial quantity, the effective energy-momentum pseudotensor $N^{\alpha\beta}$. This chapter is devoted to the analysis of this important quantity by reviewing the matter, field and harmonic gauge contributions separately. We will study these three terms $N^{\alpha\beta}_m$, $N^{\alpha\beta}_{LL}$, $N^{\alpha\beta}_H$ one after the other and extract all the relevant contributions that are within the 1.5 post-Newtonian order of accuracy.
\subsection{The effective matter pseudotensor:}
We recall from the previous chapter the precise expression for the matter contribution of the effective pseudotensor,
\begin{equation*} 
N_{m}^{\alpha\beta}\,=\,G^{-1}(\Box) \big[(-g) \ \mathcal{T}^{\alpha\beta}\big]\,=\,G(\Box)^{-1} \big[(-g) \ G(\Box) \ \mathcal{B}^{\alpha\beta}\big].
\end{equation*}
In order to extract from this expression all the relevant pieces that lie within the order of accuracy that we aim to work at in this article, we essentially need to address two different tasks. In a first step we have to review the leading terms of $\mathcal{B}^{\alpha\beta}$ (previous chapter) and see in how far they may contribute to the 1.5 post-Newtonian order of accuracy. In a second step we have to analyze how the differential operator $G^{-1}(\Box)$ acts on the product of the metric determinant $(-g)$ multiplied by the effective energy-momentum tensor $\mathcal{T}^{\alpha\beta}=G(\Box) \ \mathcal{B}^{\alpha\beta}$. Although this formal operation will lead to additional terms, the annihilation of the operator $G(\Box)$ with is inverse counterpart will substantially simplify the differential structure of the original effective energy-momentum tensor $\mathcal{T}^{\alpha\beta}$. Before we can come to the two tasks mentioned above we first need to set in place a couple of preliminary results. From a technical point of view we need to introduce the operators of instantaneous potentials \cite{Blanchet1, Blanchet3, Blanchet4},
$ \Box^{-1}[ \bar{\tau}]=\sum_{k=0}^{+\infty} \Big(\frac{\partial}{c\partial t}\Big)^{2k} \ \Delta^{-k-1}[\bar{\tau}]$. This operator is instantaneous in the sense that it does not involve any integration over time. However one should be aware that unlike the inverse retarded d'Alembert operator, this instantaneous operator will be defined only when acting on a post-Newtonian series $\bar{\tau}$. Another important computational tool which we borrow from \cite{Blanchet1, Blanchet3, Blanchet4} are the generalized iterated Poisson integrals, $\Delta^{-k-1}[\bar{\tau}_m](\textbf{x},t)=-\frac{1}{4\pi} \int d\textbf{y} \ \frac{|\textbf{x}-\textbf{y}|^{2k-1}}{2k!} \ \bar{\tau}_m(\textbf{y},t)$, where $\bar{\tau}_m$ is the $m$-th post-Newtonian coefficient of the energy-momentum source term $\bar{\tau}=\sum_{m=-2}^{+\infty} \bar{\tau}_m/c^{m}$. An additional important result that needs to be mentioned is the following generalized regularization prescription\footnote{The author would like to thank Professor E. Poisson for useful comments regarding this particular issue.}, $\big[\nabla^m \frac{1}{|\textbf{x}-\textbf{r}_A|}\big] \ \big[\nabla^n \delta(\textbf{x}-\textbf{r}_A)\big]\equiv 0, \quad \forall n,m\in \mathbb{N}$. The need for this kind of regularization prescription merely comes from the fact that inside a post-Newtonian expansion, the nonlocality of the modified Einstein equations, will lead to additional derivatives which will act on the Newtonian potentials. It is easy to see that in the limit $m=0$ and $n=0$ we recover the well known regularization prescription \cite{Poisson, Blanchet1, Blanchet2}. We are now ready to come to the first of the two tasks mentioned in the beginning of this subsection. In order to extract the pertinent pieces from $\mathcal{B}^{\alpha\beta}= \mathcal{H}(w,\Box)\ \big[\tau_m^{\alpha\beta}/(-g)\big]$ to the required order of precision, we need first to have a closer look at the differential curvature operator $\mathcal{H}(w,\Box)$. From the previous chapter we know that it is essentially composed by the potential operator function $w(h,\partial)$ and the flat spacetime d'Alembert operator,
\begin{equation*}
w(h,\partial)\,=\,-h^{\mu\nu} \partial_{\mu\nu}+\tilde{w}(h)\Box-\tilde{w}(h) h^{\mu\nu}\partial_{\mu\nu}\,=\,-\frac{h^{00}}{2}\Delta+\mathcal{O}(c^{-4}).
\end{equation*}
We see that at the 1.5 post-Newtonian order of accuracy, the potential operator function $w(h,\partial)$ reduces to one single contribution, composed by the potential $h^{00}=\mathcal{O}(c^{-2})$ \cite{Poisson, Will2, PatiWill1} and the flat spacetime Laplace operator $\Delta$. Further computational details can be found in the appendix section related to this chapter. With this in mind we can finally take up the leading four contributions of the curvature energy-momentum tensor $\mathcal{B}^{\alpha\beta}$,
\begin{equation*}
\begin{split}
\mathcal{B}^{\alpha\beta}_{1}\,&=\,\tau^{\alpha\beta}_{m}(c^{-3})-\tau^{\alpha\beta}_m(c^0) \ h^{00}+\mathcal{O}(c^{-4}),\\
B^{\alpha\beta}_{2}\,&=\,-\frac{\epsilon}{2}\sum_A m_A v^\alpha_A v^\beta_A \ \Big[\sum_{n=0}^\infty \sigma^n e^{(n+1)\kappa \Delta} \Big] \ \Big[h^{00}\Delta \delta(\textbf{y}-\textbf{r}_A)\Big]+\mathcal{O}(c^{-4}),\\
\mathcal{B}^{\alpha\beta}_3\,&=\,\frac{\epsilon\kappa}{2} e^{\kappa\Box}\Big[\frac{w^2}{1-\sigma e^{\kappa \Box}}\Big] \Big[\frac{\tau^{\alpha\beta}_m}{(-g)}\Big]\,\propto\, w^2\,=\mathcal{O}(c^{-4}),\\
\mathcal{B}^{\alpha\beta}_4\,&=\,\frac{\epsilon\kappa^2}{3!} e^{\kappa\Box}\Big[\frac{w^3}{1-\sigma e^{\kappa \Box}}\Big] \Big[\frac{\tau^{\alpha\beta}_m}{(-g)}\Big]\,\propto\, w^3\,=\mathcal{O}(c^{-6}).
\end{split}
\end{equation*}
The terms $\mathcal{B}^{\alpha\beta}_3$ and $\mathcal{B}^{\alpha\beta}_4$ are beyond the order of accuracy at which we aim to work at in this article because $\omega^2=\mathcal{O}(c^{-4})$ and $\omega^3=\mathcal{O}(c^{-6})$ and $\tau_m(c^0)$ is the matter pseudotensor at the Newtonian order of accuracy. We will see later in this chapter that  $\mathcal{B}^{\alpha\beta}_1$ will generate the usual 1.5 post-Newtonian matter source term as the second piece of the latter will precisely cancel out with another contribution. This allows us to come to the second task, namely to look at the differential operation mentioned in the introduction of this chapter,
\begin{equation*}
G^{-1}(\Box) \big[(-g) \mathcal{T}^{\alpha\beta}\big]\,=\,\big[1-\sigma e^{\kappa\Box}\big] \ \big[(-g)\mathcal{T}^{\alpha\beta}\big]
\end{equation*}
We will perform this computation using a weak-field expansion $(-g)=1+h^{00}-h^{aa}\eta_{aa}+\frac{h^2}{2}-\frac{h^{\mu\nu}h_{\mu\nu}}{4}+...$ and see how many additional terms we will produce at the 1.5 post-Newtonian order until the differential operator $G(\Box)$ and its inverse finally annihilate each other. The first term is rather simple and together with the post-Newtonian expansion for the metric determinant we obtain, 
\begin{equation*}
1 \ \big[(-g) \ \mathcal{T}^{\alpha\beta}\big]\,=\,\big[1+h^{00}\big]\mathcal{T}^{\alpha\beta}+\mathcal{O}(c^{-4}).
\end{equation*}
It should be noticed that in this relation the post-Newtonian order of $\mathcal{T}^{\alpha\beta}$ varies according to the pN-order of the quantity which it is multiplied with. The remaining contribution is by far less straightforward and needs a more careful investigation. After a rather long computation (appendix) we obtain the following result,
\begin{equation*}
-\sigma e^{\kappa \Box} \ \big[(-g) \ \mathcal{T}^{\alpha\beta}\big]\,=\,-\big[1+h^{00}\big] \big[\sigma e^{\kappa \Box}  \mathcal{T}^{\alpha\beta}\big]-\sigma D^{\alpha\beta}(c^{-3})+\mathcal{O}(c^{-4}) ,
\end{equation*}
where we have to take into account the additional tensor contribution, $D^{\alpha\beta}= \sum_{n=1}^{+\infty} \sum_{m=1}^{2n} \dbinom{2n}{m} \ \big[\nabla^{2n-m}\mathcal{T}^{\alpha\beta}\big] \big[\nabla^m h^{00}\big]$. Coming back to the initial equation for the effective matter pseudotensor $N^{\alpha\beta}_m$, we obtain by virtue of the two previous results, the following elegant expression for the modified effective matter pseudotensor,
\begin{equation*}
N_{m}^{\alpha\beta}\,=\,G^{-1}(\Box)\big[(-g) \ \mathcal{T}^{\alpha\beta}\big]\,=\,\big[1+h^{00}\big] \ \mathcal{B}^{\alpha\beta}-\sigma D^{\alpha\beta}+\mathcal{O}(c^{-4})\,=\,\mathcal{B}^{\alpha\beta}+\mathcal{B}^{\alpha\beta}h^{00}-\sigma D^{\alpha\beta}+\mathcal{O}(c^{-4}),
\end{equation*}
where we remind that $\mathcal{T}^{\alpha\beta}=G(\Box) \mathcal{B}^{\alpha\beta}$ and $\mathcal{B}^{\alpha\beta}=\mathcal{B}^{\alpha\beta}_{1}+\mathcal{B}^{\alpha\beta}_{2}+\mathcal{O}(c^{-4})$. Further computational steps are provided in the appendix related to this chapter. It is understood that there are numerous additional terms which we do not list here because they are beyond the degree of precision of this article. The two leading contributions of $N^{\alpha\beta}_m$ give rise to the usual 1.5 post-Newtonian contribution \cite{Poisson,Will2,PatiWill1,PatiWill2},
\begin{equation*}
\mathcal{B}^{\alpha\beta}_{1}(c^{-3})+\mathcal{B}_{1}^{\alpha\beta}(c^{-1})h^{00}\,=\,\sum_A m_A v_A^\alpha v^\beta_A \Big[1+\frac{\textbf{v}_A^2}{2c^2}+\frac{3V}{c^2}\Big] \ \delta(\textbf{x}-\textbf{r}_A)+\mathcal{O}(c^{-4}),
\end{equation*}
where $V=(1+\sigma) U$ is the modified Newtonian potential. The remaining task is to extract the 1.5 pN contribution out of the tensor $D^{\alpha\beta}$. We point out that this contribution is proportional to $G(\Box) \mathcal{B}^{\alpha\beta}$ as well as to the potential $h^{00}$ which is of the order $\mathcal{O}(c^{-2})$ \cite{Poisson, Will2,PatiWill1,PatiWill2}. In order to work out the contribution to the required degree of precision we need first to come back to the effective energy-momentum tensor,
\begin{equation*}
\mathcal{T}^{\alpha\beta}\,=\,G(\Box) \ \mathcal{H}(w,\Box) \ T^{\alpha\beta}\,=\,\sum_{s=0}^{+\infty} \sigma^s \sum_{p=0}^{+\infty} \frac{(s\kappa)^p}{p!} \Delta^p \bigg[\sum_A m_A v_A^\alpha v^\beta_A \ \delta(\textbf{x}-\textbf{r}_A) \bigg]+\mathcal{O}(c^{-2}),
\end{equation*}
where $G(\Box)=G(\Delta)+\mathcal{O}(c^{-2})$, $|\sigma|<1$, $\mathcal{H}(w,\Box)=1+\mathcal{O}(c^{-2})$ and $T^{\alpha\beta}=\sum_A m_A v^\alpha_A v^\beta_A \ \delta(\textbf{x}-\textbf{r}_A)+\mathcal{O}(c^{-2})$. Further computational details can be withrawn from the appendix-section related to the present chapter.
With this result at hand  we can finally write down $D^{\alpha\beta}$ to the required order of accuracy,
\begin{eqnarray*}
 D^{\alpha\beta}\,=&\,\sum_A m_A v_A^\alpha v^\beta_A \ \mathcal{S}(\sigma,\kappa) \ \Big[ \nabla^{2p+2n-m} \delta(\textbf{x}-\textbf{r}_A)\Big]\Big[ \nabla^m h^{00}\Big] +\mathcal{O}(c^{-4}).
\end{eqnarray*}
For simplicity reasons we introduced $\mathcal{S}(\sigma,\kappa)\,=\,\sum_{n=1}^{\infty} \frac{\kappa^n}{n!} \sum_{m=1}^{2n} \binom{2n}{m} \ \sum_{s=0}^{+\infty} \sigma^s \ \sum_{p=0}^{+\infty} \frac{(s\kappa)^p}{p!}$ to summarize the four sums inside $D^{\alpha\beta}$ (appendix). We remind that the first two sums come from the inverse differential operator $G^{-1}(\Box)$ while the last two sums originate from the extraction of the 1.5 pN contribution of the effective energy-momentum tensor $\mathcal{T}^{\alpha\beta}=G(\Box) \mathcal{B}^{\alpha\beta}$ and $\binom{2n}{m}=\frac{(2n)!}{(2n-m)!m!}$ is the binomial coefficient. To conclude this section we would like to point out that despite the fact that many of the contributions encountered so far contain infinitely many derivatives, we will see in the upcoming chapter that a natural post-Newtonian truncation will set in when it comes to precise computation of physical observables.
\subsection{The modified Landau-Lifshitz pseudotensor:}
In this section we will restrain our efforts to the time-time-component of the modified Landau-Lifshitz pseudotensor $N^{00}_{LL}=G^{-1}(\Box) \ \tau^{00}_{LL}$, where $\tau^{00}_{LL}=\frac{-7}{8\pi G} \partial_jV\partial^jV+\mathcal{O}(c^{-2})$ \cite{Poisson, Will2, PatiWill1}. We will see in the next chapter that this term will suffice to work out the physical quantity that we are interested in, 
\begin{eqnarray*}
c^{-2}N^{00}_{LL}\,=\, c^{-2}\Big[ \big(1-\sigma\big)\tau^{00}_{LL} -\epsilon \Delta \tau^{00}_{LL}-\sigma\sum_{m=2}\frac{\kappa^m}{m!}\Delta^m \tau^{00}_{LL}\Big]+\mathcal{O}(c^{-4}).
\end{eqnarray*}
This result was derived by using a series expansion of the exponential differential operator and by taking into account that $\partial_0=\mathcal{O}(c^{-1})$. Further computational details are provided in the appendix-section related to this chapter. The modified Landau-Lifshitz tensor contribution was scaled by the factor $c^{-2}$ for later convenience. From the leading term we will eventually be able to recover the standard post-Newtonian contribution.
\subsection{The modified harmonic gauge pseudotensor:}
The modified harmonic gauge pseudotensor contribution has the following appearance,
\begin{equation*}
N_H^{\alpha\beta}\,=\,G^{-1}(\Box) \ \tilde{\tau}^{\alpha\beta}_H\,=\,G^{-1}(\Box) \ \tau^{\alpha\beta}_H+\mathcal{O}^{\alpha\beta}.
\end{equation*}
where we remind that $\tau_H=(-g)t_H^{\alpha\beta}$ is the standard harmonic gauge pseudotensor contribution and $\mathcal{O}^{\alpha\beta}(h)=-\sigma\sum_{n=1}^{+\infty}\frac{(\kappa)^n}{n!}\partial^{2n}_0 e^{\kappa\Delta} \Box h^{\alpha\beta}$ is the iterative potential contribution. Taking into account that $h^{00}=\mathcal{O}(c^{-2})$, $h^{0a}=\mathcal{O}(c^{-3})$ and $h^{ab}=\mathcal{O}(c^{-4})$ \cite{Poisson,Will2,PatiWill1,PatiWill2} we deduce that the leading term of $\mathcal{O}^{\alpha\beta}$(h) is of the order $\mathcal{O}(c^{-4})$ or beyond, $\frac{-\epsilon}{c^{2}}  e^{\kappa\Delta} \Delta \partial_t^2h^{\alpha\beta}=\mathcal{O}(c^{-4})$. As we limit ourselves in this article to the 1.5 post-Newtonian order we do not need to consider additional correction terms coming from this contribution. It should be noticed that the higher the post-Newtonian precision is the more correction terms have to be taken into account. On the other hand we have that $\lim_{\sigma,\kappa\rightarrow 0}\mathcal{O}^{\alpha\beta}=0$ and for the same reasons mentioned in the previous subsection we will be interested in the time-time-component only. The easiest piece of the calculation by far is the computation of $\tau^{\alpha\beta}_H=(-g)t^{\alpha\beta}_H$ to the required degree of accuracy. Using the results from \cite{Poisson, Will2, PatiWill1} for the purely general relativistic harmonic gauge contribution we can easily deduce that, $\frac{16\pi G}{c^4} (-g)t^{00}_H=\mathcal{O}(c^{-6})$, is beyond the order of accuracy that we are interested in in this article. It is straightforward to see that the same is true for the modified pseudotensor $\frac{16\pi G}{c^4} N^{00}_H\,=\,\mathcal{O}(c^{-6})$. 
\section{The effective total mass:}
The total near zone mass of a N-body system \cite{Poisson, Will2, PatiWill1, PatiWill2} is composed by the matter and the field energy confined in the region of space defined by $\mathcal{M}:|\textbf{x}|<\mathcal{R}$, $M=c^{-2} \int_\mathcal{M} d\textbf{x} \ \big(N^{00}_m+N^{00}_{LL}\big)\,=\,M_m+M_{LL}+\mathcal{O}(c^{-4})$. The modified harmonic gauge contribution, $N_H=\mathcal{O}(c^{-4})$, is beyond the 1.5 post-Newtonian order of accuracy. The matter and field contributions will be worked out separately before they will be combined to form the effective total near zone mass. We saw in the previous chapter that the matter contribution can be rephrased in a more detailed way by splitting up the modified matter pseudotensor $N^{00}_m$ into its different components. This partition will eventually allow to review the different contributions one after the other and to retain all the terms that are within desired degree of accuracy, 
\begin{equation*}
M_m\,=\,c^{-2}\int_{\mathcal{M}}d\textbf{x}\ N^{00}_m\,=\,c^{-2}\int_{\mathcal{M}}d\textbf{x} \ \big[\mathcal{B}^{00}+\mathcal{B}^{00}h^{00}-\sigma D^{00}\big]+\mathcal{O}(c^{-4}).
\end{equation*}
We will start our investigation by analysing a piece that will essentially lead to the general relativistic 1.5 pN term $M_m^{GR}$ \cite{Poisson}.
\begin{equation*}
\begin{split}
M_{\mathcal{B}_1+\mathcal{B}_1h^{00}}=&\,c^{-2}\int_{\mathcal{M}} d \textbf{x} \ \big[\mathcal{B}_1^{00}+\mathcal{B}_1^{00}h^{00}\big]\,=\,M_m^{GR}+3\sigma\frac{G}{c^2} \sum_A\sum_{B \neq A} \frac{m_A m_B}{ r_{AB}}+\mathcal{O}(c^{-4}),
\end{split}
\end{equation*}
where $r_{AB}=|\textbf{r}_A-\textbf{r}_B|$ is the distance between body $A$ and body $B$. It should be noticed that the second term in this expression, which could have been presented in a more succinct way by simply writing $M^{NL}_M$, merely originates from the modified Newtonian potential introduced in the third chapter of this article. The next contribution of the effective curvature energy-momentum tensor $\mathcal{B}^{\alpha\beta}=\mathcal{H}(\Box,\omega) T^{\alpha\beta}$ which could potentially contribute to the effective total near zone mass, at the 1.5 pN order of accuracy, is the $\mathcal{B}^{\alpha\beta}_2$ piece. A careful analysis (appendix) however reveals that, at this order of accuracy, it cannot contribute to the total mass because of its high order nonlocal structure, $M_{\mathcal{B}_2}=c^{-2}\int_{\mathcal{M}} d \textbf{x} \ \mathcal{B}_2^{00}=0$. Indeed after multiple partial integration the differential operator acting on the potential $h^{00}$ is of order two or higher, so that we will encounter surface terms and terms proportional to, $\sum_A\sum_{B\neq A} m_Am_B \  \nabla^m \delta(\textbf{r}_A-\textbf{r}_B)\,=\,0, \quad \forall m\in \mathbb{N}$ only. It can be easily seen, by a having a look at its differential operator structure, that the same reasoning is true for the derivative term $D^{\alpha\beta}$ encountered for the first time in the previous chapter, $M_{D}=c^{-2}\int_{\mathcal{M}} d \textbf{x} \ D^{00}=0$. In both cases surface terms can be freely discarded as we limit ourselves in this article only to the near zone domain, $\int_{\partial \mathcal{M}} dS^p \ \partial_p h^{00}\frac{\delta(\mathbf{y}-\mathbf{r}_A)}{|\mathbf{x}-\mathbf{y}|}  \propto \delta(\mathcal{R}-|\mathbf{r}_A|)=0$.  Additional inside on the derivation of this and the previous result can be found in the appendix related to this chapter. The remaining series of terms belonging to the effective matter pseudotensor $N^{00}_m$ are beyond the 1.5 post-Newtonian order of accuracy. Summing up all the non-vanishing terms we finally obtain the total effective matter contribution,
\begin{equation*}
\begin{split}
M_m\,=M_m^{GR}+M_m^{NL}+\mathcal{O}(c^{-4}),
\end{split}
\end{equation*}
where we introduced for clarity reasons the following two independent mass-terms, $M_m^{GR}=\sum_Am_A+\frac{1}{2c^2}\sum_A m_A v^2_A+3\frac{G}{c^2}\sum_A\sum_{B\neq A} \frac{m_Am_B}{r_{AB}}$ and $M_m^{NL}=3\sigma\frac{G}{c^2}\sum_A\sum_{B\neq A} \frac{m_Am_B}{r_{AB}}$. Here $M_m^{GR}$ is the standard general relativistic term at the 1.5 post-Newtonian order of accuracy \cite{Poisson} and $M_m^{NL}$ is the additional contribution originating from the nonlocal coupling operator $G(\Box_g)$, worked out to the same order of accuracy. It is straightforward to observe that in the limit of vanishing modification parameters ($\sigma,\kappa\rightarrow 0$) this result gently reduces to the general relativistic one. It was seen in the previous chapters that the field contribution of the total effective mass is obtained by evaluating the following integral,
\begin{equation*}
\begin{split}
M_{LL}\,=\,c^{-2}\int_{\mathcal{M}} d\textbf{x} \ N^{00}_{LL}\,= \,c^{-2}\int_{\mathcal{M}} d\textbf{x} \ \big[ (1-\sigma)\tau^{00}_{LL}-\epsilon \Delta \tau^{00}_{LL} -\sigma \sum_{m=2}^{+\infty} \frac{\kappa^m}{m!} \Delta^m \tau^{00}_{LL}\big]+\mathcal{O}(c^{-4}),
\end{split}
\end{equation*}
where we remind the important result $\tau^{00}_{LL}=\frac{-7}{8\pi G} \partial_pV\partial^pV$ and  $V=(1+\sigma) \ U$ is the effective Newtonian potential introduced in chapter three. We will review the three different contributions one after the other and study in how far they will eventually contribute to the total effective  gravitational near zone mass. The first integral gives essentially rise to the usual 1.5 pN general relativistic Landau-Lifshitz field-term \cite{Poisson},
\begin{equation*}
\begin{split}
c^{-2}\int_{\mathcal{M}} d\textbf{x} \ \tau^{00}_{LL}\,=\,-\frac{7G}{2c^2} \sum_A\sum_{B\neq A} \frac{ \tilde{m}_A \tilde{m}_B}{|\textbf{r}_A-\textbf{r}_B|}\,=\,(1+\sigma)^2 M_{LL}^{GR},
\end{split}
\end{equation*}
where we recall that $\tilde{m}_A=(1+\sigma) \ m_A$ is the effective mass of body $A$. We refer the reader interested in the precise derivation of this result to the appendix related to this chapter. The remaining two terms do not contribute for the same reasons that were outlined before when we investigated a possible 1.5 pN contribution from the $\mathcal{B}^{\alpha\beta}_2$ and $D^{\alpha\beta}$ terms,
 \begin{equation*}
\begin{split}
\frac{\epsilon}{c^2}\int_{\mathcal{M}} d\textbf{x} \  \Delta \tau_{LL}^{00}\,=\,0,\quad \ \frac{\sigma}{c^2}\int_{\mathcal{M}} d\textbf{x} \  \sum_{m=2}^{+\infty} \frac{\kappa^m}{m!} \Delta^m \tau^{00}_{LL}\,=\,0.
\end{split}
\end{equation*}
We provide additional computational details about the precise derivation of these two results in the appendix relative to this chapter. Strictly speaking these two terms give rise to $\mathcal{R}$-dependent terms. However they can be discarded as they will cancel out with their wave zone counterparts as was shown in \cite{PatiWill1}. In analogy to the previous subsection, we intend to terminate the present one by providing the total near zone field (Landau-Lifshitz) mass at the 1,5 pN order of precision,  
\begin{equation*}
M_{LL}\,=\,M_{LL}^{GR}+M_{LL}^{NL}+\mathcal{O}(c^{-4}).
\end{equation*}
We distinguish between the standard general relativistic piece $M_{LL}^{GR}$ \cite{Poisson} and the the additional contribution originating from the nonlocal coupling operator,
\begin{equation*}
\begin{split}
M_{LL}^{GR}\,=\,-\frac{7G}{2c^2} \sum_A\sum_{B\neq A}  \frac{m_A m_B}{|\textbf{r}_A-\textbf{r}_B|},\quad M_{LL}^{NL}\,=\,-v(\sigma)\frac{7G}{2c^2} \sum_A\sum_{B\neq A}  \frac{m_A m_B}{|\textbf{r}_A-\textbf{r}_B|}.
\end{split}
\end{equation*}
We see that the nonlocal contribution is of 1.0 pN order, $M_{LL}^{NL}=\mathcal{O}(c^{-2})$. In the limit of vanishing $\sigma$ the nonlocal field term disappears as the polynomial, $v(\sigma)=\sigma-\sigma^2-\sigma^3$, depends only on the dimensionless parameter $\sigma$. We obtain, after joining the matter and field the contributions, the total gravitational near zone mass,
\begin{equation*}
M=M^{GR}+M^{NL}+\mathcal{O}(c^{-4}),
\end{equation*}
where we have have introduced the following two quantities, $M^{GR}=M_m^{GR}+M_{LL}^{GR}$ and $M^{NL}=M_m^{NL}+M_{LL}^{NL}$ in order to distinguish between the standard general relativistic terms and the nonlocal contributions,
\begin{equation*}
\begin{split}
M^{GR}\,=\,\sum_Am_A+\frac{1}{c^2}\sum_A \frac{m_A v^2_A}{2}-\frac{1}{2}\frac{G}{c^2}\sum_A\sum_{B\neq A} \frac{m_Am_B}{r_{AB}},\quad M^{NL}\,=\,z(\sigma)\frac{G}{c^2}\sum_A\sum_{B\neq A} \frac{m_Am_B}{r_{AB}}.
\end{split}
\end{equation*}
The newly introduced function $z(\sigma)$ is another polynomial of the modification parameter $\sigma$: $z(\sigma)=3\sigma-\frac{7}{2}v(\sigma)=-\sigma/2 +7/2 \ (\sigma^2+\sigma^3)$. It is obvious from what has been said previously that we recover the usual 1.5 PN general relativistic near-zone mass in the limit of vanishing $\sigma$.
\section{Conclusion:}
In this article we outlined a precise model of a nonlocally modified theory of gravity in which Newton's constant $G$ is promoted to a differential operator $G_\Lambda(\Box_g)$. Although the nonlocal equations of motion are themselves generally covariant, they cannot (for nontrivial $G_\Lambda(\Box_g)$) be presented as a metric variational derivative of a diffeomorphism invariant action unless you assume that they are only a first, linear in the curvature, approximation for the complete equations of motion \cite{Barvinsky1, Barvinsky2}. The general idea of a differential coupling was apparently formulated for the first time in \cite{Dvali1,Barvinsky1,Dvali2,Barvinsky2} in order to address the cosmological constant problem \cite{Weinberg1}. However the idea of a varying coupling constant of gravitation dates back to early works of Dirac \cite{Dirac1} and Jordan \cite{Jordan1, Jordan2}. Inspired by these considerations Brans and Dicke published in the early sixties a theory in which the gravitational constant is replaced by the reciprocal of a scalar field \cite{Brans1}. We presented the general idea of infrared degravitation in which $G_\Lambda(\Box_g)$ acts like a high-pass filter with a macroscopic distance filter scale $\sqrt{\Lambda}$. In this way sources characterized by characteristic wavelengths much smaller than the filter scale  ($\lambda_c\ll\sqrt{\Lambda}$) pass (almost) undisturbed through the filter and gravitate normally, whereas sources characterized by wavelengths larger than the filter scale are effectively filtered out \cite{Dvali1,Dvali2}. We concluded chapter one by reviewing the cosmological constant problem and outlined a precise differential coupling model by which we can observe an effective degravitation of the vacuum energy on cosmological scales. In the second chapter we worked out the relaxed Einstein equations in the context of ordinary gravity and we briefly introduced the post-Newtonian theory as well as related concepts that were used in the subsequent chapters. In chapter three we derived the effective relaxed Einstein equations and showed that in the limit of vanishing UV parameters and infinitely large IR parameter we recover the standard wave equation. In analogy to the purely general relativistic case we worked out a formal near-zone solution for a far away wave-zone field point in terms of the effective energy-momentum pseudotensor $N^{\alpha\beta}$. The latter forms the main body of chapter four in which we worked out separately its matter, field and harmonic gauge contributions ($N^{\alpha\beta}_m$, $N^{\alpha\beta}_{LL}$, $N^{\alpha\beta}_H$) up to the 1.5 post-Newtonian order of accuracy. In the penultimate chapter the previous results were gathered in order to work out the effective total 1.5 post-Newtonian near-zone mass. We observe that in the limit of vanishing UV parameters we recover the standard 1.5 post-Newtonian total near-zone mass. 
\begin{acknowledgments}
The author would like to thank Professor Eric Poisson (University of Guelph) for useful comments regarding the generalized regularization prescription. A. D. gratefully acknowledges support by the Ministry for Higher Education and Research of the G.-D. of Luxembourg (MESR-Cedies).
\end{acknowledgments}

\appendix
\section{The relaxed Einstein equations:}
We consider a material source consisting of a collection of fluid balls \cite{Poisson, Will2,PatiWill1,PatiWill2} whose size is typically small compared to their separations, $T^{\alpha \beta}=\rho\ u^\alpha u^\beta$, where $\rho=\frac{\rho^*}{\sqrt{-g}\gamma_A}$ is the energy-density and $u^\alpha=\gamma_A (c,\textbf{v}_A)$ is the four-velocity of the fluid ball with point mass $m_A$ and individual trajectory $r_A(t)$. Taking into account that for point masses we have $\rho^*=\sum_{A=1}^Nm_A\ \delta\big(\textbf{x}-\textbf{r}_A(t)\big)$, $\frac{1}{\sqrt{-g}}=1-\frac{1}{2}h^{00}+\mathcal{O}(c^{-4})$ and $\gamma_A^{-1}=\sqrt{-g_{\mu\nu}\frac{v^\mu_Av^\nu_A}{c^2}}=1-\frac{1}{2}\frac{\textbf{v}_A^2}{c^2}-\frac{1}{4}h^{00}+\mathcal{O}(c^{-4})$, we obtain the 1.5 post-Newtonian matter energy-momentum pseudotensor outlined in chapter two, $c^{-2}(-g)T^{00}=(1+\frac{4U}{c^2})\sum_Am_A\delta(\textbf{x}-\textbf{r}_A)(1+\frac{v_A^2}{c^2}-\frac{U}{c^2})+\mathcal{O}(c^{-4})$. Bearing in mind that $h^{00}=\mathcal{O}(c^{-2})$, $h^{0a}=\mathcal{O}(c^{-3})$, $h^{ab}=\mathcal{O}(c^{-4})$ and that $\partial_0h^{00}$ is of order $c^{-1}$ relative to $\partial_ah^{00}$ we see that the dominant piece of $\tau^{\alpha\beta}_{LL}$ will come from $\partial_ah^{00}=\frac{4}{c^2}\partial_a U$, where we remind that $U$ is the Newtonian potential of the N-body system. Moreover each occurrence of $g_{\alpha\beta}$ can be replaced by $\eta_{\alpha\beta}$ because each factor of $h^{\alpha\beta}$ contributes a power of $G$ and we aim to compute $\tau^{\alpha\beta}_{LL}$ to order $G^2$ in the second post-Minkowskian approximation. Using $\mathfrak{g}^{\alpha\beta}=\eta^{\alpha\beta}-h^{\alpha\beta}$ and the harmonic gauge $\partial_\beta h^{\alpha\beta}=0$, we can review the remaining six contributions of the time-time component of the Landau-Lifshitz pseudotensor $\tau_{LL}^{00}$ presented in chapter two,
\begin{equation*}
\begin{split}
\frac{1}{2}g^{00}g_{\lambda\mu}\partial_\rho \mathfrak{g}^{\lambda\nu}\partial_\nu\mathfrak{g}^{\mu\rho}\,=&\,-\frac{1}{2}\eta_{\lambda\mu}\partial_\rho h^{\lambda\nu}\partial_\nu h^{\mu\rho}=\mathcal{O}(c^{-6}),\\
-g^{0\lambda}g_{\mu\nu}\partial_\rho\mathfrak{g}^{0\nu}\partial_\lambda\mathfrak{g}^{\mu\rho}\,=&\,-\eta^{0\lambda}\eta_{\mu\nu}\partial_\rho h^{0\nu}\partial_\lambda h^{\mu\rho}\,=\,\mathcal{O}(c^{-6}),\\
-g^{0\lambda}g_{\mu\nu}\partial_\rho\mathfrak{g}^{0\nu}\partial_\lambda\mathfrak{g}^{\mu\rho}\,=&\,-\eta^{0\lambda}\eta_{\mu\nu}\partial_\rho h^{0\nu}\partial_\lambda h^{\mu\rho}\,=\,\mathcal{O}(c^{-6}),\\
g_{\lambda\mu}g^{\nu\rho}\partial_\nu\mathfrak{g}^{0\lambda}\partial_\rho\mathfrak{g}^{0\mu}\,=&\,-\partial^bh^{00}\partial_b h^{00}+\mathcal{O}(c^{-6}),\\
\frac{1}{4}(2g^{0\lambda}g^{0\mu}-g^{00}g^{\lambda\mu})g_{\nu\rho}g_{\sigma\tau}\partial_\lambda\mathfrak{g}^{\nu\tau}\partial_\mu\mathfrak{g}^{\rho\sigma}\,=&\,\frac{1}{4}\partial^bh^{00}\partial_bh^{00}+\mathcal{O}(c^{-6}),\\
-\frac{1}{8}(2g^{0\lambda}g^{0\mu}-g^{00}g^{\lambda\mu})g_{\rho\sigma}g_{\nu\tau}\partial_\lambda\mathfrak{g}^{\nu\tau}\partial_\mu\mathfrak{g}^{\rho\sigma}\,=&\,-\frac{1}{8}\partial^bh^{00}\partial_bh^{00}+\mathcal{O}(c^{-6}).
\end{split}
\end{equation*} 
Summing up all terms that make up $\tau^{00}_{LL}$ we finally obtain, $\frac{16 \pi G}{c^4}(-g)t^{00}_{LL}=-\frac{7}{8}\partial_bh^{00}\partial^bh^{00}+\mathcal{O}(c^{-6})$.
\section{The modified relaxed Einstein equations:}
\subsection{The effective energy-momentum tensor:}
We have for an arbitrary contravariant rank two tensor $f^{\alpha\beta}(x)$ \cite{Poisson, Weinberg2, Woodard1, Maggiore2},
\begin{equation*}
\begin{split}
\Box_g f^{\alpha\beta}(x)\,
%=&\,\nabla^\mu \nabla_\mu f^{\alpha\beta}(x)\\
%&=\,\frac{1}{\sqrt{-g}}\partial_\mu\big(\sqrt{-g}g^{\mu\nu}\partial_\nu f^{\alpha\beta}(x)\big)\\
=\,\frac{1}{\sqrt{-g}}\partial_\mu\Big[(\eta^{\mu\nu}-h^{\mu\nu})\partial_\nu f^{\alpha\beta}(x)\Big]\,
%=&\,\frac{1}{\sqrt{-g}} \Big[\eta^{\mu\nu}\partial_\mu \partial_\nu f^{\alpha\beta}(x)-\partial_\mu h^{\mu\nu} \partial_\nu f^{\alpha\beta}(x)-h^{\mu\nu}\partial_\mu \partial_\nu f^{\alpha\beta}(x)\Big]\\
&=\,\frac{1}{\sqrt{-g}}\Big[\Box f^{\alpha\beta}(x)-h^{\mu\nu}\partial_\mu\partial_\nu f^{\alpha\beta}(x)\Big]\\
&=\,\Big[1-\frac{h}{2}+\frac{h^2}{8}-\frac{h^{\rho\sigma}h_{\rho\sigma}}{4}+\mathcal{O}(G^3)\Big]^{-1}\Big[\Box f^{\alpha\beta}(x)-h^{\mu\nu}\partial_\mu\partial_\nu f^{\alpha\beta}(x)\Big]\\
%=&\,\Big[1+\frac{h}{2}-\frac{h^2}{8}+\frac{h^{\rho\sigma}h_{\rho\sigma}}{4}+\mathcal{O}(G^3)\Big]\Big[\Box f^{\alpha\beta}(x)-h^{\mu\nu}\partial_\mu \partial_\nu f^{\alpha\beta}(x)\Big]\\
%=&\,\Box f^{\alpha\beta}(x) -h^{\mu\nu}\partial_{\mu\nu}f^{\alpha\beta}(x)\\
%&+\,\Big[\frac{h}{2}-\frac{h^2}{8}+\frac{h^{\rho\sigma}h_{\rho\sigma}}{4}+\mathcal{O}(G^3)\Big] \Box f^{\alpha\beta}(x)\\
%&-\,\Big[\frac{h}{2}-\frac{h^2}{8}+\frac{h^{\rho\sigma}h_{\rho\sigma}}{4}+\mathcal{O}(G^3)\Big]h^{\mu\nu}\partial_{\mu} \partial_\nu f^{\alpha\beta}(x)\\
&=\, \big[\Box-h^{\mu\nu}\partial_\mu\partial_\nu+\tilde{w}(h)\Box-\tilde{w}(h)h^{\mu\nu}\partial_{\mu}\partial_\nu\big] f^{\alpha\beta}(x)\\
%&=\, \big[\Box+w(h,\partial)\big]f^{\alpha\beta}(x),
\end{split}
\end{equation*}
where the harmonic gauge conditions $\partial_\mu h^{\mu\nu}\,=\,0$ were used together with $\sqrt{-g}g^{\mu\nu}=\eta^{\mu\nu}-h^{\mu\nu}$ \cite{Poisson, Will2,PatiWill1,PatiWill2, Blanchet1} and the following definition $\tilde{w}(h)= \frac{h}{2}-\frac{h^2}{8}+\frac{h^{\rho\sigma}h_{\rho\sigma}}{4}+\mathcal{O}(G^3)$ was introduced for the potential function. In the quest of decomposing the effective energy-momentum tensor $\mathcal{T}^{\alpha\beta}=G(\Box) \ \mathcal{H}(w,\Box) \ T^{\alpha\beta}$ we should remind the important result for linear differential operators, $[A,B]=0\Rightarrow [\frac{1}{A},\frac{1}{B}]=0$, where $A$ and $B$ are supposed to be two linear differential operators. Let $f$ be a function with $f\in \mathcal{C}^{\infty}(\mathbb{R})$.
We have that $ABf=BAf\Leftrightarrow BA^{-1}g=A^{-1}Bg\Leftrightarrow A^{-1}B^{-1}h=B^{-1}A^{-1}h$, where $f=A^{-1}g$ and $g=B^{-1}h$ and therefore $g$, $h$ $\in \mathcal{C}^{\infty}(\mathbb{R})$.
This result will be used in the splitting of the effective energy-momentum tensor,
\begin{equation*}
\begin{split}
\mathcal{T}^{\alpha\beta}\,=\, G\big[\Box_g\big] \ T^{\alpha\beta}
 %&=\, \frac{1}{1-\sigma e^{\kappa\Box_g}}  \ T^{\alpha\beta}\\
&=\, \frac{1}{1-\sigma e^{\kappa \Box}} \ \Big[1-\sigma \frac{e^{\kappa \Box}}{1-\sigma e^{\kappa \Box}}\sum_{n=1}^\infty \frac{\kappa^n}{n!} w^n\Big]^{-1} \  \ T^{\alpha\beta}\\
 &=\,\Big[\frac{1}{1-\sigma e^{\kappa \Box}} \ \Big(1+\sigma \frac{e^{\kappa \Box}}{1-\sigma e^{\kappa \Box}} \sum_{n=1}^\infty \frac{\kappa^n}{n!} w^n+\mathcal{O}(\sigma^2)\Big)\Big]  \ T^{\alpha\beta}\\
&=\,G\big[\Box\big] \ \Big[\sum_{n=0}^{+\infty}\mathcal{B}^{\alpha\beta}_n+\mathcal{O}(\sigma^2)\Big],
\end{split}
\end{equation*}
where we used $1-\sigma e^{\kappa\Box_g}=1-\sigma e^{\kappa[\Box+\omega(h,\partial)]}=1-\sigma e^{\kappa\Box}\ \sum_{n=0}^{+\infty} \frac{\kappa^n}{n!}\omega^n$. Moreover we needed to constrain the range for the modulus of the dimensionless parameter $\sigma$ which, from now on, has to be smaller than one in order to make the perturbative expansion work. We adopt the convention that differential operators appearing in the numerator act first ($[w,\Box]\neq 0$). For linear differential operators this prescription is not needed as they commute anyway. The leading four contributions of the curvature tensor are displayed in the main part of the article.
\subsection{The modified relaxed Einstein Equations:}
In the present appendix-subsection we would like to present some additional computational details regarding the modified Green function outlined in the main part. By substituting the Fourier representation of the modified Green function $G(x-y)=(2\pi)^{-1}\int dk\ G(k) e^{ik(x-y)}$, where $x=(ct,\textbf{x})$ and $k=(k^0,\textbf{k})$, inside Green the function condition $(1-\sigma e^{\kappa \Delta})\Box G(x-y)=\delta(x-y)$ we obtain the latter in momentum-space, $G(k)= \frac{1}{(k^0)^2-|\textbf{k}|^2} \ \frac{1}{1-\sigma e^{-\kappa \textbf{k}^2}}
= \frac{1}{(k^0)^2-|\textbf{k}|^2}+\sigma \ \frac{ \ e^{-\kappa|\textbf{k}|^2}}{(k^0)^2-|\textbf{k}|^2}+...$ The first term in this infinite expansion is the usual Green function followed by correction terms. We also remind that the modulus of the dimensionless parameter is assumed to be strictly smaller than one, $|\sigma|<1$. By making use of the residue theorem we can derive the modified Green function, $G=G^{GR}+G^{NL}$, in terms of its retarded and advanced contributions, $G^{GR}=\frac{-1}{4\pi} \frac{1}{|\textbf{x}-\textbf{y}|} \Big[\delta(x^0-|\textbf{x}-\textbf{y}|-y^0)-\delta(x^0+|\textbf{x}-\textbf{y}|-y^0)\Big]$ and $G^{NL}=\frac{-1}{4\pi} \frac{1}{|\textbf{x}-\textbf{y}|} \frac{\sigma}{2\sqrt{\kappa \pi}}\Big[ e^{-\frac{x^0-|\textbf{x}-\textbf{y}|-y^0}{4\kappa}}-e^{-\frac{x^0+|\textbf{x}-\textbf{y}|-y^0}{4\kappa}}\Big]$. The modified Newtonian potential is obtained from the formal solution of the modified relaxed Einstein equation, $h^{00}=\frac{4G}{c^4}\int dy \ G_r(x-y) N^{00}(x)+\mathcal{O}(c^{-1})=\frac{4G}{c^4}\int d\textbf{y}\int dy^0\ \Big[\delta(x^0-|\textbf{x}-\textbf{y}|-y^0)+\frac{\sigma}{2\sqrt{\kappa \pi}}e^{-\frac{(x^0-|\textbf{x}-\textbf{y}|-y^0)^2}{4\kappa}}\Big]\frac{\sum_A m_A c^{2} \delta(\textbf{y}-\textbf{r}_A)}{|\textbf{x}-\textbf{y}|}+\mathcal{O}(c^{-1})$. After performing the four dimensional integration we recover the result for the time-time component of the gravitational potential outlined in the main text of this article.
\subsection{Solution for a far way wave-zone field point:}
In analogy to \cite{Poisson, Will2, PatiWill1, PatiWill2}, we aim to expand the retarded effective pseudotensor in terms of a power series ,
\begin{equation*}
\begin{split}
 \frac{N^{\alpha\beta}(x^0-|\textbf{x}-\textbf{y}|,\textbf{y})}{|\textbf{x}-\textbf{y}|}\,&=\, \sum_{l=0}^{\infty} \frac{(-1)^l}{l!} \textbf{y}^L\partial_L \Big[\frac{N^{\alpha\beta}(x^0-r,\textbf{y})}{r}\Big]\\
&=\, \frac{N^{\alpha\beta}}{r}-y^a \frac{\partial}{\partial x^a} \big[\frac{N^{\alpha\beta}}{r}\big]+\frac{y^ay^b}{2} \frac{\partial^2}{\partial x^a\partial x^b} \big[\frac{N^{\alpha\beta}}{r}\big]-\cdots\\
%&=\,\frac{1}{r} \ \sum_{l=0}^\infty \frac{(-1)^l}{l!} y^L\partial_L \ N^{\alpha\beta}(u,\textbf{y})+\mathcal{O}(1/r^2)\\
&=\,\frac{1}{r} \ \sum_{l=0}^\infty \frac{y^L}{l!}  \ n_L \ \Big(\frac{\partial}{\partial u}\Big)^l \ N^{\alpha\beta}(u,\textbf{y})+\mathcal{O}(1/r^2),
%&=\,\frac{1}{r} \ \sum_{l=0}^\infty \frac{1}{l!}\frac{1}{c^l} y^L \ n_L \ \Big(\frac{\partial}{\partial t_r}\Big)^l \ N^{\alpha\beta}(u,\textbf{y})+\mathcal{O}(1/r^2)
\end{split}
\end{equation*}
where we used the following result,
\begin{equation*}
\begin{split}
\partial_L N^{\alpha\beta}\,=\, \frac{\partial}{\partial x^{a1}} \ \cdots \ \frac{\partial}{\partial x^{al}} N^{\alpha\beta}
%&=\, \frac{\partial}{\partial x^{a1}} \ \cdots \ \frac{\partial N^{\alpha\beta}}{\partial u} \frac{\partial u}{\partial x^{al}}\\
\,=\, \Big(\frac{\partial}{\partial u}\Big)^l \ N^{\alpha\beta} \ \frac{\partial u }{\partial x^{a1}} \ \cdots \ \frac{\partial u}{\partial x^{al}}
%&=\, \Big(\frac{\partial}{\partial u}\Big)^l \ N^{\alpha\beta} \ \frac{\partial (x^0-r) }{\partial x^{a1}} \ \cdots \ \frac{\partial (x^0-r)}{\partial x^{al}}\\
%&=\, \Big(\frac{\partial}{\partial u}\Big)^l \ N^{\alpha\beta} \ (-1)^l \ n_{a1} \ \cdots \ n_{al}\\
\,=\, (-1)^l \ \Big(\frac{\partial}{\partial u}\Big)^l \ N^{\alpha\beta}  \ n_{L}, 
\end{split}
\end{equation*}
and where $\frac{\partial r}{\partial x^a}=\frac{x^a}{r}=n_a$ is the a-th componant of the radial unit vector and $u=c\tau=x^0-r$. The far away wave zone is characterized by the fact that we only need to consider the potentials contribution proportional to $1/r$. This allows us to derive the near-zone contribution of the gravitational potentials for a wave-zone field point in terms of the retarded derivatives,
\begin{equation*}
\begin{split}
h^{ab}_{\mathcal{N}}(x)\,=\, \frac{4 G}{c^4} \sum_{l=0}^\infty \frac{(-l)^l}{l!} \partial_L \Big[\frac{1}{r} \int_{\mathcal{M}} d\textbf{y} \ N^{ab}(u,\textbf{y}) \ y^L \Big]\,&=\, \frac{4 G}{c^4 r} \sum_{l=0}^\infty \frac{n_L}{c^ll!} \Big(\frac{\partial}{\partial \tau}\Big)^l \Big[ \int_{\mathcal{M}} d\textbf{y} \ N^{ab}(u,\textbf{y}) \ y^L\Big]+\mathcal{O}(r^{-2})\\
&=\,\frac{4 G}{c^4 r}\Bigg[\int_{\mathcal{M}} d\textbf{y} \ N^{ab}(u,\textbf{y}) + \frac{n_c }{c} \frac{\partial}{\partial \tau} \int_{\mathcal{M}} d\textbf{y} \ N^{ab}(u,\textbf{y}) \ y^c\\
&\quad \ +\frac{n_c n_d}{2c^2} \frac{\partial^2}{\partial \tau^2} \int_{\mathcal{M}} d\textbf{y} \ N^{ab}(u,\textbf{y}) \ y^c \ y^d\\
&\quad \ +\frac{n_c n_d n_e}{6c^3} \frac{\partial^3}{\partial \tau^3} \int_{\mathcal{M}} d\textbf{y} \ N^{ab}(u,\textbf{y}) \ y^c \ y^d \ y^e+[l\ge4]\Bigg]+\mathcal{O}(r^{-2})
\end{split}
\end{equation*}
We provide some additional details on how the modified conservation relations were derived for an arbitrary domain of integration $\mathcal{M}$ with boundary $\partial \mathcal{M}$,
\begin{equation*}
\begin{split}
\partial^2_0 \int_{\mathcal{M}} d\textbf{x} \ N^{00} \ x^ax^b\,&=\,\partial_0\Big[\int_{\mathcal{M}} d\textbf{x} \ \Big(N^{0a}x^b+N^{0b}x^a-\partial_c(N^{0c}x^ax^b)\Big)\Big]\\
&=\,\int_{\mathcal{M}} d\textbf{x} \ \Big(2N^{ab}+\partial_c(\partial_dN^{dc}x^ax^b)\Big)-\int_{\partial\mathcal{M}} dS_c\ \big(N^{ca}x^b+N^{cb}x^a\big)\\
&=\,\int_\mathcal{M} d\textbf{x} \ \Big(2N^{ab}-\partial_c(N^{ca}x^b+N^{cb}x^a-\partial_dN^{dc}x^ax^b\Big).
\end{split}
\end{equation*}
Additional details on the derivation of the second conservation identity,
\begin{equation*}
\begin{split}
\partial_0\int_{\mathcal{M}} d\textbf{x} \ \Big(N^{0a}x^bx^c+N^{0b} x^ax^c -N^{0c} x^ax^b\Big)\,=&\,\int_{\mathcal{M}} d\textbf{x} \Big(-\partial_dN^{da} x^bx^c-\partial_dN^{db}x^ax^c+\partial_dN^{dc} x^ax^b\Big)\\
=&\,\int_{\mathcal{M}} d\textbf{x} \ \Big(2N^{ab} x^c-\partial_d(N^{da}x^bx^c+N^{db}x^ax^c-N^{dc}x^ax^b)\Big).
\end{split}
\end{equation*}
This and the previous result have been worked out by making use of the conservation relation ($\partial_\beta N^{0\beta}=0$), (multiple) partial integration and the Gauss-Ostrogradsky-theorem. Furthermore it should be pointed out that we can easily replace the derivative $\partial_0$ by $\partial u$ as the two variables only differ by a constant shift in time. The precise form of the surface terms, mentioned in the main part of the article, is,
\begin{equation*}
P^{ab}\,=\, \int_{\partial \mathcal{M}} dS_c \ (N^{ac} y^b+N^{bc} y^a-\partial_d N^{cd} y^a y^b), \ 
P^{abc}\,=\, \frac{1}{c}\frac{\partial}{\partial \tau} \int_{\partial \mathcal{M}} dS_d \ (N^{ad} y^b y^c+N^{bd} y^a y^c -N^{cd} y^a y^b).
\end{equation*}
\section{The effective energy-momentum pseudotensor:}

\subsection{The effective matter pseudotensor:}
Taking into account that $h^{00}=\mathcal{O}(c^{-2})$, $h^{0a}=\mathcal{O}(c^{-3})$ and $h^{ab}=\mathcal{O}(c^{-4})$ \cite{Poisson, Will2, PatiWill1}, we see that the three contributions of the potential operator function $w(h,\partial)$ are of the following post-Newtonian orders ($h=\eta_{\alpha\beta}h^{\alpha\beta}$), $h^{\mu\nu}\partial_{\mu\nu}=\mathcal{O}(c^{-4})$, $\tilde{w}(h)=\frac{h}{2}-\frac{h^2}{8}+\frac{h^{\rho\sigma}h_{\rho\sigma}}{4}+\mathcal{O}(G^3)=-\frac{h^{00}}{2}+\mathcal{O}(c^{-4})$, $\tilde{w}(h)h^{\mu\nu}\partial_{\mu\nu}=\mathcal{O}(c^{-6})$. The leading contribution of $\mathcal{B}^{\alpha\beta}$ gives rise to the usual 1.5 post-Newtonian matter contribution,
\begin{equation*}
\begin{split}
\mathcal{B}^{\alpha\beta}_1\,=\,\frac{\tau^{\alpha\beta}_m}{(-g)}\,&=\,\Big[\tau^{\alpha\beta}_m(c^{-3})+\mathcal{O}(c^{-4})\Big]\Big[1-h^{00}+h^{aa}-\frac{h^2}{2}+\cdots\Big]\,=\,\tau^{\alpha\beta}_{m}(c^{-3})-\tau^{\alpha\beta}_m(c^0) \ h^{00}+\mathcal{O}(c^{-4}),
\end{split}
\end{equation*}
where $\tau^{\alpha\beta}_m$ is the effective matter pseudotensor introduced in chapter two.
The second contribution of $\mathcal{B}^{\alpha\beta}$ is more advanced and can be decomposed at the 1.5 pN order of accuracy into three different contributions,
\begin{equation*}
\begin{split}
\mathcal{B}^{\alpha\beta}_2\,&=\,\epsilon e^{\kappa \Box}\Big[\frac{w}{1-\sigma e^{\kappa\Box}}\Big] \  \Big[\frac{\tau_m^{\alpha\beta}}{(-g)}\Big]\,
%&=\,-\frac{\epsilon}{2} \sum_A m_A v_A^\alpha v^\beta_A \ \sum_{n=0}^{+\infty} \sigma^n\Bigg[1+[n+1]\kappa \Delta+\sum_{m=2}^{+\infty} \frac{[(n+1)\kappa]^m}{m!} \Delta^m\Bigg] \\
%&\quad\quad \Bigg[h^{oo} \Big(\Delta \delta(\textbf{y}-\textbf{r}_A)\Big)\Bigg]+\mathcal{O}(c^{-4})\\
=\,-\frac{\epsilon}{2}\sum_A m_A v^\alpha_A v^\beta_A \ \Big[\sum_{n=0}^\infty \sigma^n e^{(n+1)\kappa \Delta} \Big] \ \Big[h^{00}\Delta \delta(\textbf{y}-\textbf{r}_A)\Big]+\mathcal{O}(c^{-4}),
\end{split}
\end{equation*}
where we used $\tau_m^{\alpha\beta}=\sum_A m_A v_A^\alpha v_A^\beta \delta(\textbf{y}-\textbf{r}_A)+\mathcal{O}(c^{-2})$ with $(-g)=1+h^{00}+\mathcal{O}(c^{-4})$ and $\omega=-\frac{h^{00}}{2}+\mathcal{O}(c^{-4})$. For later purposes we will decompose this quantity into three different pieces, $\mathcal{B}^{\alpha\beta}_2= \mathcal{B}_{2a}+\mathcal{B}_{2b}+\mathcal{B}_{2c}+\mathcal{O}(c^{-4})$ where,
\begin{equation*}
\begin{split}
\mathcal{B}_{2a}^{\alpha\beta}\,=&\,-\frac{\epsilon}{2} \frac{1}{1-\sigma} \sum_A m_A v_A^\alpha v^\beta_A \  \Big[h^{00} \big(\Delta \delta(\textbf{y}-\textbf{r}_A)\big)\Big], \ \quad \mathcal{B}_{2b}^{\alpha\beta}\,=\, -\frac{\epsilon}{2} \frac{\kappa}{(1-\sigma)^2} \sum_A m_A v_A^\alpha v^\beta_A \Delta \Big[h^{00} \big(\Delta \delta(\textbf{y}-\textbf{r}_A)\big)\Big],\\ \mathcal{B}_{2c}^{\alpha\beta}\,= &\,-\frac{\epsilon}{2} \sum_A m_A v_A^\alpha v^\beta_A \   \sum_{n=0}^{+\infty} \sigma^n \ \sum_{m=2}^{+\infty} \frac{[(n+1)\kappa]^m}{m!} \Delta^m \Big[h^{00} \big(\Delta \delta(\textbf{y}-\textbf{r}_A)\big)\Big].
\end{split}
\end{equation*} 
This splitting was obtained using the definition \cite{Spallucci1} for an exponential differential operator $e^{(n+1)\kappa\Delta}=1+[n+1]\kappa \Delta+\sum_{m=2}^{+\infty} \frac{[(n+1)\kappa]^m}{m!} \Delta^m$ and we assumed that the modulus of the dimensionless parameter is smaller than one $|\sigma|<1$, so that we have, $\sum_{n=0}^{+\infty} \sigma^n=\frac{1}{1-\sigma}$ and $\sum_{n=0}^{+\infty} [n+1]\sigma^n=\frac{1}{(1-\sigma)^2}$. To work out the 1.5 pN terms originating from the exponential differential operator acting on the product of the effective energy-momentum tensor and the metric determinant, we have to rely on the generalized Leibniz product rule $\forall\ q(x), \ v(x) \in \mathcal{C}^\infty(\mathbb{R}),\quad \big(q(x)v(x)\big)^{(n)}=\sum_{k=0}^n\binom{n}{k} q^{(k)} v^{(n-k)}$, where $\binom{n}{k}=\frac{n!}{k!(n-k)!}$ are the binomial coefficients,
\begin{equation*}
\begin{split}
\sigma e^{\kappa \Box} \Big[\mathcal{T}^{\alpha\beta} (-g)\Big]
&\,=\, \sigma e^{-\kappa\partial^2_{0}} e^{\kappa \Delta} \Big[\mathcal{T}^{\alpha\beta} (-g)\Big]\\
&=\, \sigma \Big[\sum_{s=0}^{\infty} \frac{(-\kappa)^s}{s!} \Big(\partial^2_0\Big)^s\Big] \Big[\sum_{n=0}^{\infty} \frac{(\kappa)^n}{n!} \Big(\nabla^2\Big)^n\Big] \Big[\mathcal{T}^{\alpha\beta} (-g)\Big]\\
&=\, \sigma \sum_{s=0}^{\infty} \frac{(-\kappa)^s}{s!}\sum_{n=0}^{\infty} \frac{\kappa^n}{n!} \sum_{m=0}^{2n} \binom{2n}{m} \sum_{p=0}^{2s} \binom{2s}{p}\Big[\partial^{2s-p}_0\Big(\nabla^{2n-m} \mathcal{T}^{\alpha\beta}\Big)\Big] \Big[\partial_0^p\Big(\nabla^m (-g)\Big)\Big]\\
&=\,\sigma \sum_{s=0}^{\infty} \frac{(-\kappa)^s}{s!}\sum_{n=0}^{\infty} \frac{\kappa^n}{n!} \binom{2n}{0} \binom{2s}{0} \Big[\partial_0^{2s}\Big(\nabla^{2n}\mathcal{T}^{\alpha\beta}\Big)\Big](-g)+\\
&\quad \ \,\sigma \sum_{s=1}^{\infty} \frac{(-\kappa)^s}{s!}\sum_{n=0}^{\infty} \frac{\kappa^n}{n!} \sum_{p=1}^{2s} \binom{2n}{0} \binom{2s}{p} \Big[\partial_0^{2s-p}\Big(\nabla^{2n} \mathcal{T}^{\alpha\beta}\Big)\Big] \Big[\partial_0^p (-g)\Big]+\\
&\quad \ \,\sigma \sum_{s=0}^{\infty} \frac{(-\kappa)^s}{s!}\sum_{n=1}^{\infty} \frac{\kappa^n}{n!} \sum_{m=1}^{2n} \binom{2n}{m} \binom{2s}{0}\Big[\partial_0^{2s} \Big(\nabla^{2n-m} \mathcal{T}^{\alpha\beta}\Big) \Big] \Big[\nabla^m (-g)\Big]+\\
&\quad \ \,\sigma \sum_{s=1}^{\infty} \frac{(-\kappa)^s}{s!}\sum_{n=1}^{\infty} \frac{\kappa^n}{n!} \sum_{m=1}^{2n} \binom{2n}{m}\sum^{2s}_{p=1} \binom{2s}{p} \Big[\partial_0^{2s-p} \nabla^{2n-m} \mathcal{T}^{\alpha\beta}\Big] \Big[\partial_0^p\Big(\nabla^m(-g)\Big)\Big]\\
%&=\, \Big[\sigma e^{\kappa \Box} \mathcal{T}^{\alpha\beta}\Big] \Big(1+f(h)\Big)+\\
%&\,\sigma \frac{(-\kappa)^1}{1!} \sum_{p=1}^2 C^{2}_{p} \Big[\partial_0^{2-p} \Big(e^{\kappa \Delta} \mathcal{T}^{\alpha\beta}\Big)\Big] \Big[\partial_0^p f(h)\Big]+\\
%&\, \sigma \sum_{s=2}^{\infty}\frac{(-\kappa)^s}{s!} \sum_{p=1}^{2s} C^{2s}_{p} \Big[\partial_0^{2s-p} \Big(e^{\kappa \Delta} \mathcal{T}^{\alpha\beta}\Big)\Big] \Big[\partial_0^p f(h)\Big]+\\
%&\, \sigma \sum_{n=1}^{\infty} \frac{\kappa^n}{n!} \sum_{m=1}^{2n} C^{2n}_{m} \Big[ e^{-\kappa \partial_0^2} \Big(\nabla^{2n-m} \mathcal{T}^{\alpha\beta}\Big) \Big] \Big[ \nabla^m f(h)\Big]-\\
%&\,\sigma\kappa \sum_{n=1}^{\infty} \frac{\kappa^n}{n!} \sum_{m=1}^{2n} \sum_{p=1}^{2} C^{2n}_{m} C^{2}_{p} \Big[\partial_0^{2-p} \nabla^{2n-m} \mathcal{T}^{\alpha\beta}\Big] \Big[\partial_0^p \Big(\nabla^m f(h)\Big)\Big]+\\
%&\,\sigma \sum_{s=2}^{\infty} \frac{(-\kappa)^s}{s!}\sum_{n=1}^{\infty} \frac{\kappa^n}{n!} \sum_{m=1}^{2n} C^{2n}_{m} \sum^{2s}_{p=1} C^{2s}_{p} \Big[\partial_0^{2s-p} \nabla^{2n-m} \mathcal{T}^{\alpha\beta}\Big] \Big[\partial_0^p\Big(\nabla^m f(h)\Big)\Big]\\
&=\,\sigma [1+h^{00}] \ e^{\kappa\Box}\mathcal{T}^{\alpha\beta}+ \sigma \sum_{n=1}^{\infty} \frac{\kappa^n}{n!} \sum_{m=1}^{2n} \binom{2n}{m} \Big[ \Big(\nabla^{2n-m} \mathcal{T}^{\alpha\beta}\Big) \Big] \Big[ \nabla^m h^{00}\Big]+\mathcal{O}(c^{-4}).
\end{split}
\end{equation*}
We remind that $(-g)=1+h^{00}+\mathcal{O}(c^{-4})$, $\partial_0=\mathcal{O}(c^{-1})$ and $\binom{2n}{0}=\binom{2s}{0}=1$. 
In order to compute $D^{\alpha\beta}= \sum_{n=1}^{\infty} \frac{\kappa^n}{n!} \sum_{m=1}^{2n} \binom{2n}{m} \big[ \big(\nabla^{2n-m} \mathcal{T}^{\alpha\beta}\big) \big]\big[\nabla^m h^{00}\big]$ to the required order of accuracy wee need to work out the effective energy-momentum tensor to lowest post-Newtonian order $\mathcal{T}^{\alpha\beta}= G(\Box) \ \mathcal{H}(w,\Box) \ T^{\alpha\beta}$, where $G(\Box)=\sum_{s=0}^{+\infty} \sigma^s e^{s\kappa\Box}\,=\,\sum_{s=0}^{+\infty}\sigma^s\sum_{p=0}^{+\infty}\frac{(s\kappa\Delta)^p}{p!}+\mathcal{O}(c^{-2})$ and $\mathcal{H}(\omega,\Box)=1+\sum_{s=0}^{+\infty}\sigma^{s+1}\sum_{p=0}^{+\infty}\frac{\big((s+1)\kappa\Box\big)^p}{p!}\sum_{n=1}^{+\infty}\frac{\kappa^n}{n!} \omega^n+\cdots=1+\mathcal{O}(c^{-2})$. We remind that we assumed $|\sigma|<1$ and we have $\Box=\Delta+\mathcal{O}(c^{-2})$ and the potential operator function $\omega(h,\partial)=\mathcal{O}(c^{-2})$. We would like to conclude this appendix section by recalling the energy-momentum tensor of a system composed by N particles with negligible pressure, $T^{\alpha\beta}=\rho \ u^\alpha u^\beta$, where $\rho=\frac{\rho^*}{\sqrt{-g} \gamma_A}$, $\rho^*=\sum_{A=1}^Nm_A\delta\big(\textbf{x}-\textbf{r}_A(t)\big)$, $\gamma_A^{-1}=\sqrt{-g_{\mu\nu}\frac{v^\mu_Av^\nu_A}{c^2}}=1-\frac{1}{2}\frac{\textbf{v}_A^2}{c^2}-\frac{1}{4}h^{00}+\mathcal{O}(c^{-4})$, $u^\alpha=\gamma_A (c,\textbf{v}_A)$ is the relativistic velocity \cite{Poisson, Will2, PatiWill1, PatiWill2} and $\textbf{r}_A(t)$ is the individual trajectory of the particle with mass $m_A$.
\subsection{The effective Landau-Lifshitz pseudotensor:}
From the series expansion of the exponential differential operator \cite{Spallucci1} we obtain $G^{-1}(\Box)\tau^{00}_{LL}=\big[1-\sigma \sum_{m=0}^{+\infty}\frac{(\kappa\Box)^n}{n!}\big]\tau^{00}_{LL}$.
\section{The effective total mass:}
\subsection{Matter contribution $M_m$:}
We provide additional computational steps in order to show how the results, presented in the main text, have been derived. In this appendix we will focus mainly on technical issues. We therefore refer the reader to the explanations given in the main text for the notations and conceptual points. The leading order matter contribution is,
\begin{equation*}
\begin{split}
M_{\mathcal{B}_1+\mathcal{B}_1h^{00}}\,=&\,c^{-2}\int_{\mathcal{M}} d \textbf{x} \ \Big[\mathcal{B}_1^{00}+\mathcal{B}^{00}h^{00}\Big]\\
=&\,c^{-2}\int_{\mathcal{M}} d \textbf{x} \ \sum_A m_A c^2 \Big[1+\frac{v_A^2}{2c^2}+3(1+\sigma)\frac{G}{c^2}\sum_{B\neq A}\frac{m_B}{|\textbf{x}-\textbf{r}_B|}\Big]\delta(\textbf{x}-\textbf{r}_A)+\mathcal{O}(c^{-4})\\
=&\,M_m^{GR}+3\sigma\frac{G}{c^2} \sum_A\sum_{B \neq A} \frac{m_A m_B}{ r_{AB}}+\mathcal{O}(c^{-4}),
\end{split}
\end{equation*}
where we used the standard regularization prescription $\frac{\delta(\textbf{x}-\textbf{r}_A)}{|\textbf{x}-\textbf{r}_A|}\equiv 0$ for point masses \cite{Poisson, Blanchet1, Blanchet2}.
We would like to illustrate why $M_{\mathcal{B}_{2}}=0$. In order to do so we isolate the lowest derivative term in $\mathcal{B}_{2}$ presented in chapter four and its related appendix-section, $M_{\mathcal{B}_{2}}=-\frac{\epsilon}{2} \frac{1}{1-\sigma} \sum_A m_A  \int_{\mathcal{M}}d\textbf{x} \ h^{00} \Big[\Delta \delta(\textbf{y}-\textbf{r}_A)\Big]+\cdots$.
In what follows we will show that this term cannot contribute to the total mass,
\begin{equation*}
\begin{split}
 \sum_A m_A  \int_{\mathcal{M}}d\textbf{x} \ h^{00} \Big[\Delta \delta(\textbf{x}-\textbf{r}_A)\Big]\,=&\,\sum_A m_AS_1-\sum_A m_AS_2-4\pi \frac{4G}{c^2} \sum_A \sum_{B\neq A} m_A\tilde{m}_B  \int_{\mathcal{M}}d\textbf{x} \ \delta(\textbf{x}-\textbf{r}_B) \ \delta(\textbf{x}-\textbf{r}_A)\\
=&\,-4\pi\frac{4G}{c^2} \sum_A \sum_{B\neq A} m_A\tilde{m}_B \ \delta(\textbf{r}_B-\textbf{r}_A)\,=\,0,
\end{split}
\end{equation*}
where $S_1=\oint_{\partial\mathcal{M}}dS^ph^{00}[\partial_p\delta(\textbf{x}-\textbf{r}_A)]$ and $S_2=\oint_{\partial\mathcal{M}}dS^p[\partial_ph^{00}]\delta(\textbf{x}-\textbf{r}_A)$ are the surface integrals originating from partial integration. We remind that the time-time-component of the gravitational potential for a N-body system is $h^{00}=\frac{4}{c^2} V=(1+\sigma)\frac{4}{c^2}\sum_{B} \frac{m_B}{|\textbf{x}-\textbf{r}_B|}=\frac{4}{c^2}\sum_{B} \frac{\tilde{m}_B}{|\textbf{x}-\textbf{r}_B|}$ and we used the well known identity, $\Delta \frac{1}{|\textbf{x}-\textbf{r}_B|}=-4\pi \delta(\textbf{x}-\textbf{r}_B)$. Surface terms of the form, $\oint_{\partial \mathcal{M}} dS^p \ \partial_p h^{00}\delta(\mathbf{x}-\mathbf{r}_A)  \propto \delta(\mathcal{R}-|\mathbf{r}_A|)=0$, coming from partial integration, vanish in the near zone defined by $\mathcal{M}: \ |\textbf{x}|<\mathcal{R}$ \cite{Poisson}, For the higher order derivative terms in $\mathcal{B}_2$ the situation is very similar in the sense that we will always encounter, after (multiple) partial integration, terms of the form, $\sum_A\sum_{B\neq A} m_Am_B \ \int_{\mathcal{M}} d\textbf{x} \  \delta(\textbf{x}-\textbf{r}_A) \ \nabla^m \delta(\textbf{x}-\textbf{r}_B)=0, \quad \forall m\in \mathbb{N}$. Very similar arguments show that we have, at this order of accuracy, for the contribution $M_D=c^{-2}\int_\mathcal{M}D^{00}=\sum_Am_A\mathcal{S}(\sigma,\kappa)[\nabla^{2p+2n-m}\delta(\textbf{x}-\textbf{r}_A)][\nabla^mh^{00}]=0$, where we remind that $\mathcal{S}(\sigma,\kappa)=\sum_{n=1}^\infty\frac{\kappa^n}{n!}\sum_{m=1}^{2n}\dbinom{2n}{m}\sum_{s=0}^{+\infty}\sigma^s\sum_{p=0}^{+\infty}\frac{(s\kappa)^p}{p!}$ is the four-sum introduced in the previous chapter.
\subsection{Field contribution $M_{LL}$:}
Here we will have a closer look at the important term,
\begin{equation*}
\begin{split}
c^{-2}\int_{\mathcal{M}} d\textbf{x} \ \tau^{00}_{LL}\,=\,-\frac{7}{8c^{2}\pi G}\int_{\mathcal{M}} d\textbf{x} \  \partial_pV\partial^pV\,=&\,-\frac{7}{8c^2\pi G}\int_{\mathcal{M}} d\textbf{x} \  \Big[\partial_p(V\partial^pV)-V\nabla^2V\Big]\\
=&\,-\frac{7}{8c^2\pi G}\int_{\mathcal{M}} d\textbf{x} \  \Big[\partial_p(V\partial^pV)+4\pi G \sum_A\tilde{m}_A\delta(\textbf{x}-\textbf{r}_A)V\Big]\\
=&\,f_a(\mathcal{R})-\frac{7G}{2c^2} \sum_A\sum_{B\neq A} \frac{ \tilde{m}_A \tilde{m}_B}{|\textbf{r}_A-\textbf{r}_B|},
\end{split}
\end{equation*}
where we remind that $V=(1+\sigma)\ U$ is the effective Newtonian potential and $\tilde{m}_A=(1+\sigma)\ m_A$ is the effective mass of body $A$.
By virtue of the Gauss-Ostrogradsky theorem the first term gives rise to an $\mathcal{R}$-dependent contribution which will eventually cancel out with the corresponding wave zone term \cite{Poisson, Will2, PatiWill1},
\begin{equation*}
\begin{split}
-\frac{8c^{2}\pi G}{7}f_a(\mathcal{R})\,=\,\int_\mathcal{M}d\textbf{x}\ \partial_p(V\partial^pV)\,=\,\oint_{\partial_\mathcal{M}}dS_p(V\partial^pV)\,=&\,\sum_{A,B}\tilde{m}_A\tilde{m}_B\oint_{\partial\mathcal{M}}dS^p\frac{G^2}{|\textbf{x}-\textbf{r}_A|}\frac{(\textbf{r}_B-\textbf{x})_p}{|\textbf{x}-\textbf{r}_B|^3}\\
=&\,\sum_{A,B}\tilde{m}_A\tilde{m}_B\frac{G^2}{\mathcal{R}}\int N^pN_p\ d\Omega+\mathcal{O}(r_{AB}/\mathcal{R})\\
=&\,4\pi\sum_{A,B}\tilde{m}_A\tilde{m}_B\frac{G^2}{\mathcal{R}}+\mathcal{O}(r_{AB}/\mathcal{R}).
\end{split}
\end{equation*}
According to \cite{Poisson} this integral can be evaluated by using the substitution $\textbf{y}=\textbf{x}-\textbf{r}_B$, so that $\textbf{x}-\textbf{r}_A=\textbf{y}-\textbf{r}_{AB}$, where $\textbf{r}_{AB}=\textbf{r}_A-\textbf{r}_B$ is the relative separation between body $A$ and $B$, $\textbf{N}=\textbf{y}/y$ is the surface element on the boundary defined by $y=\mathcal{R}$ and $dS^p=\mathcal{R}^2 N^p d\Omega$. We used the fact that the relative distance between the bodies is much smaller than the scale of the near zone domain, $\frac{1}{|\textbf{y}-\textbf{r}_{AB}|}\big|_{|\textbf{y}|=\mathcal{R}}=\frac{1}{\mathcal{R}}+\mathcal{O}(r_{AB}/\mathcal{R})$, where $r_{AB}=|\textbf{r}_{AB}|$ as well as $\int N^pN_p\ d\Omega=4\pi$ and $d\Omega=\sin\theta d\theta d\phi$ is an element of solid angle in the direction specified by $\theta$ and $\phi$ \cite{Poisson}. The higher order derivative contributions are,
\begin{equation*}
\begin{split}
\frac{\epsilon}{c^2}\int_{\mathcal{M}} d\textbf{x} \  \Delta \tau_{LL}^{00}[c^{-3}]\,=&\,f_b(\mathcal{R})-\epsilon\frac{7}{2}\frac{G}{c^2} \sum_A\sum_{B\neq A} \tilde{m}_A \tilde{m}_B \ \int_{\mathcal{M}} d\textbf{x} \ \Delta \bigg[  \frac{\delta(\textbf{x}-\textbf{r}_A)}{|\textbf{x}-\textbf{r}_B|}\bigg]\,=\,f_b(\mathcal{R})\\
\frac{\sigma}{c^2}\int_{\mathcal{M}} d\textbf{x} \  \sum_{m=2}^{+\infty} \frac{\kappa^m}{m!} \Delta^m \tau^{00}_{LL}[c^{-3}]\,=&\, f_c(\mathcal{R})-\sigma\frac{7}{2} \frac{G}{c^2} \sum_A\sum_{B\neq A} \tilde{m}_A \tilde{m}_B \  \sum_{m=2}^{+\infty} \frac{\kappa^m}{m!}\int_{\mathcal{M}} d\textbf{x} \ \Delta^m  \bigg[\frac{\delta(\textbf{x}-\textbf{r}_A)}{|\textbf{x}-\textbf{r}_B|}\bigg]\,=\,f_c(\mathcal{R}).
\end{split}
\end{equation*}
The last two results have been derived by using the relation, $\partial_pV\partial^pV=\partial_p(V\partial^pV)+4\pi G\sum_A\tilde{m}_A\delta(\textbf{x}-\textbf{r}_A)V$ ($V=(1+\sigma) \ U$). The second term in the first of the two integrals above,
\begin{equation*}
\begin{split}
\sum_A \tilde{m}_A \int_\mathcal{M}d\textbf{x}\ \Delta [\delta(\textbf{x}-\textbf{r}_A)h^{00}]\,=\,\sum_A \tilde{m}_AS_1+\sum_A \tilde{m}_AS_2+(2-2)4\pi\frac{4G}{c^2} \sum_A \sum_{B\neq A} \tilde{m}_A\tilde{m}_B \ \delta(\textbf{r}_B-\textbf{r}_A)\,=\,0,
\end{split}
\end{equation*}
vanishes after double partial integration. The surface integrals $S_1$ and $S_2$ were defined in the context of the computation of $M_{\mathcal{B}_2}$. Multiple partial integration was used and surface terms, being proportional to $\delta(\mathcal{R}-\textbf{r}_A)$, were discarded as they do not contribute to the near zone $\mathcal{M}: \ |\textbf{x}|<\mathcal{R}$. Discarding all the remaining $\mathcal{R}$-depending terms, $f_i(\mathcal{R})$ with $i\in \{a,b,c\}$, for the same reasons that have been mentioned above, we finally obtain the results given in the main chapter \cite{Poisson, Will2, PatiWill1}.

\end{document}